\documentclass{article}
\usepackage{microtype}
\usepackage{graphicx}
\usepackage{booktabs} 
\usepackage[skip=0pt]{parskip}
\usepackage{pdfpages}
\usepackage{hyperref}
\usepackage{mathtools}
\usepackage{amsmath}
\usepackage{cite}
\usepackage{amsmath,amsfonts}
\usepackage{graphicx}
\usepackage{textcomp}
\usepackage{xcolor}
\usepackage{mathtools}
\usepackage{xspace}
\usepackage{microtype}
\usepackage{graphicx}
\usepackage{subcaption}

\usepackage{sidecap, caption}
\usepackage{booktabs} 
\usepackage{hyperref}
\hypersetup{ 
    colorlinks=true, 
    linkcolor=blue,
    anchorcolor=black,
    citecolor=purple,
    filecolor=cyan,
    menucolor=red,
    runcolor=cyan,
    urlcolor=magenta
} 
\usepackage{multirow}
\usepackage{bm}
\usepackage{apptools}
\usepackage{graphicx}
\usepackage{hyperref}
\usepackage{eso-pic}
\usepackage{xparse}
\usepackage{xspace}
\usepackage[ruled,noline,noend,linesnumbered]{algorithm2e}
\usepackage{amsmath}
\usepackage{enumitem}
\usepackage{comment}
\usepackage[accepted]{mlsys2025}

\newcommand{\name}{GSplit\xspace}

\newcommand{\Tname}{Split parallelism\xspace}
\newcommand{\tname}{split parallelism\xspace}
\newcommand{\Tadj}{Split-parallel\xspace}

\newcommand{\tadj}{split-parallel\xspace}
\newcommand{\ds}{mixed frontier\xspace}

\newcommand{\mypar}[1]{\vspace*{0.05in}\noindent\textbf{#1.}}

\newcommand{\sandremoved}[1]{{}}

\DeclareMathOperator*{\E}{\mathbb{E}}
\mlsystitlerunning{\name: Scaling Graph Neural Network Training on Large Graphs via Probabilistic Splitting}

\newlength{\badgewidth}
\newlength{\badgegap}
\newcommand{\badgeList}{}

\NewDocumentCommand{\addTopRightBadge}{O{} m}{%
\gappto{\badgeList}{\href{#1}{\includegraphics[width=\badgewidth]{#2}}\hspace{\badgegap}}%
}

\newcommand{\placeTopRightBadges}{%
\AddToShipoutPictureBG*{%
\put(\LenToUnit{\paperwidth - 1.5cm - \badgewidth},\LenToUnit{\paperheight - 2cm}){%
\makebox[0pt][r]{\badgeList}%
}%
}%
}

\addTopRightBadge{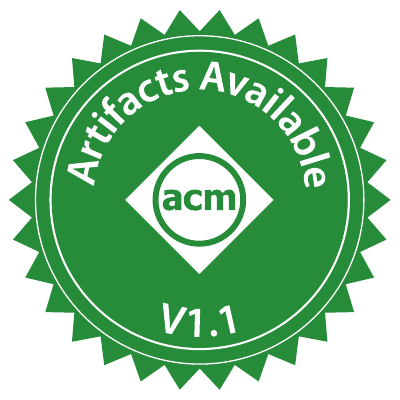}
\addTopRightBadge{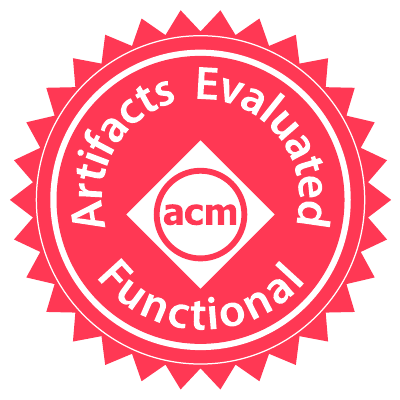}

\placeTopRightBadges
\begin{document}

\twocolumn[
\mlsystitle{\name: Scaling Graph Neural Network Training on Large Graphs via Split-Parallelism}
\mlsyssetsymbol{equal}{*}

\begin{mlsysauthorlist}
\mlsysauthor{Sandeep Polisetty}{equal,umass}
\mlsysauthor{Juelin Liu}{equal,umass}
\mlsysauthor{Jacob Falus}{umass}
\mlsysauthor{Yi Ren Fung}{uiuc}
\mlsysauthor{Seung Hwan Lim}{ornl}
\mlsysauthor{Hui Guan}{umass}
\mlsysauthor{Marco Serafini}{umass}
\end{mlsysauthorlist}

\mlsysaffiliation{umass}{University of Massachusetts, Amherst}
\mlsysaffiliation{ornl}{Oak Ridge National Laboratory}
\mlsysaffiliation{uiuc}{University of Illinois Urbana-Champaign }

\mlsyscorrespondingauthor{Sandeep Polisetty}{spolisetty@umass.edu}

\mlsyskeywords{Machine Learning, MLSys}

\vskip 0.3in

\begin{abstract}
    Graph neural networks (GNNs), an emerging class of machine learning models for graphs, have gained popularity for their superior performance in various graph analytical tasks. 
    Mini-batch training is commonly used to train GNNs on large graphs, and data parallelism is the standard approach to scale mini-batch training across multiple GPUs. 
    Data parallel approaches contain redundant work as subgraphs sampled by different GPUs contain significant overlap.
    To address this issue, we introduce a hybrid parallel mini-batch training paradigm called \tname. 
    \Tname avoids redundant work by splitting the sampling, loading, and training of each mini-batch across multiple GPUs. 
    \Tname, however, introduces communication overheads that can be more than the savings from removing redundant work.
    We further present a lightweight partitioning algorithm that probabilistically minimizes these overheads.
    We implement \tname in \name and show that it outperforms state-of-the-art mini-batch training systems like DGL, Quiver, and $P^3$.
\end{abstract}
]
\printAffiliationsAndNotice{

\mlsysEqualContribution
}
\section{Introduction}
Graph neural networks (GNNs) have demonstrated superior performance in various graph analytics tasks. 
Widely used systems like DGL~\citep{dgl} and PyTorch Geometric~\citep{pytorch-geometric}, as well as production systems like AliGraph~\citep{aligraph}, employ mini-batch training, which is generally more effective than full-graph training at scaling to multiple GPUs~\cite{bajaj2024graphneuralnetworktraining}. 
To accelerate GNN model training across multiple GPUs, these systems utilize \emph{data parallelism}. 
At each iteration, they sample multiple independent \textit{micro-batches}, one per GPU. 
Each GPU independently loads its micro-batch and computes gradients using a local replica of the GNN model.
Each micro-batch consists of a partition of the target vertices in the mini-batch along with a sample of their k-hop neighbors.

\begin{figure*}
    \centering
    \includegraphics[width=0.95 \textwidth]{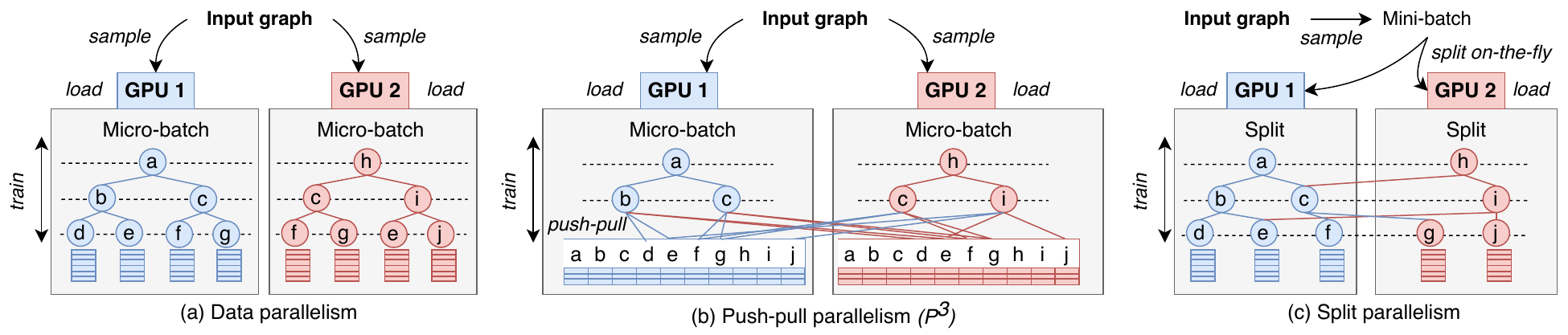}
    \caption{Comparison between data parallel, push-pull parallel, and the proposed split parallel training.}
    \label{fig:comparison}
\end{figure*}

One drawback of data-parallel training for GNNs is redundant data movements and computations (see Figure~\ref{fig:comparison}(a)).
The k-hop neighbors of target vertices in different micro-batches overlap. Thus, the same vertices appear in multiple micro-batches. 
This redundancy leads to overheads across all stages of the graph neural network model training pipeline -- sampling, loading, and training.

In this paper, we propose using a different hybrid parallelism approach tailored to mini-batch training that eliminates redundant data loading and computation.
During each training iteration, the sampling step samples one mini-batch for all GPUs, divides it on the fly into non-overlapping partitions called \emph{splits}, and assigns each split to a specific GPU (see Figure~\ref{fig:comparison}(c)).
Now, only one GPU is responsible for the sampling, feature loading, and computation steps associated with a vertex.
GPUs then cooperatively execute the training of an iteration on the same mini-batch. 
Each GPU operates only on the vertices within its assigned split. It shuffles intermediate results with other GPUs at each GNN layer. 
We refer to this parallelism technique as \emph{\tname}. 
We implement \tname in \name, a scalable multi-GPU GNN training system.
\name delivers state-of-the-art performance in mini-batch GNN training by eliminating redundancy.
However, applying \tname to mini-batch GNN training requires solving several challenges.

The primary technical challenge in split parallelism is devising a \emph{splitting algorithm} that satisfies three key requirements: (i) minimize the cost of shuffles while balancing load to avoid stragglers (ii) on-the-fly splitting at each iteration without becoming a performance bottleneck, (iii) Split vertices in alignment with the location of their cached input features on the GPU.
A straightforward solution to the first requirement would be to run a min-cut graph partitioning algorithm online on each sampled mini-batch.
However, this would not satisfy the second requirement as the partitioning problem is NP-hard and difficult to parallelize, nor the third, since it would not consider where input features are cached.
Many full-graph training systems shuffle intermediate data among GPUs at each GNN layer and use sophisticated scheduling algorithms to minimize communication and balance load~\citep{roc, neugraph, cai2021dgcl, md2021distgnn, betty, pipegcn, mgg, G3}.
However, these algorithms do not apply directly to splitting.
In full-graph training, the work performed in each epoch is known in advance, since the batch is static and training occurs on the entire graph.
In contrast, a splitting algorithm must dynamically split sampled mini-batches at each iteration, which operates on a much shorter time scale than a full-graph epoch.

To solve this problem, we propose a \emph{probabilistic splitting algorithm} that achieves a negligible online overhead: on a randomly sampled mini-batch, it \emph{provably} minimizes the \emph{expected} communication cost among splits and balances the \emph{expected} load per split.
Compared to using a standard offline graph partitioning algorithm that does not provide probabilistic guarantees, our algorithm speeds up the end-to-end training time of our \name system by up to 1.7$\times$.

Another challenge is to preserve the programming abstractions of data-parallel training systems.
Besides simplifying development, preserving the programming abstractions of data-parallel training allows us to leverage work on optimizing single-GPU kernels for GNN sampling and training~\citep{adaptgear, fastkernel, wu2021seastar, nextdoor, gsampler, fu2022tlpgnn, ye2023sparsetir}.
In data-parallelism, vertices in a micro-batch layer are always local, whereas in \tname a layer may contain a mix of local and remote vertices.
Further, \name provides a \emph{split/shuffle} API that hides the low-level data shuffling details from the end users.

Our evaluation across multiple large graphs and GNN models shows that \name outperforms the state-of-the-art systems like DGL~\citep{dgl} and Quiver~\citep{quiver} by up to 4.4$\times$ and 1.9$\times$ respectively (2.4$\times$ and 1.4$\times$ on average). 
We also implement and evaluate the push-pull parallelism approach of $P^3$~\citep{gandhi2021p3} in a single-host multi-GPU system and show that \name outperforms it by up to 4.1$\times$ (2.4$\times$ on average).
\name's splitting algorithm is the key to achieving these speedups, and it is effective in balancing load among splits and reducing the cost of shuffles.

Overall, we make the following contributions:
\begin{itemize}[noitemsep,nosep]
    \sandremoved{
    \item We characterize the cost of data loading in mini-batch GNN training, including also $P^3$'s push-pull parallelism (Section~\ref{sec:problem}). 
    }
    \item We propose \tname to eliminate redundant input feature loading and computation (Section~\ref{sec:opportunities}) \sandremoved{ and discuss how we embodied it in the \name system (Section~\ref{sec:overview}). 
    }
    \item We propose a lightweight online splitting algorithm that uses a probabilistic approach to minimize the expected communication cost and balance the expected load per split at each iteration (Section~\ref{sec:workload}).
    \item We develop \name's split/shuffle API, which supports optimized single-GPU kernels for sampling and training (Section~\ref{sec:api}).   
\end{itemize}
\section{Background and Motivation}
\label{sec:problem}
This section first introduces mini-batch GNN training, and then elaborates on the limitations of existing optimizations. 

\begin{figure}[t]
    \centering
\includegraphics[width=0.48\textwidth]{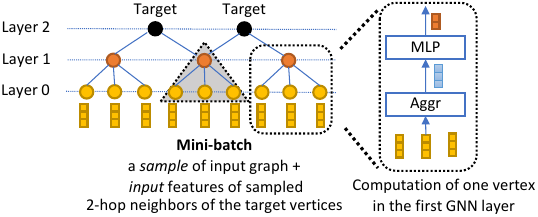}
    \caption{Example of a mini-batch.}
    \label{fig:minibatch}
    \vspace{-15pt}
\end{figure}

\textbf{Mini-Batch GNN Training and Data Parallelism} 
A GNN model is defined as a sequence of \emph{GNN layers}.
During each mini-batch training iteration, there are three phases: \textit{sampling}, \textit{loading}, and \textit{training}.
Given a set of target vertices, the sampling phase randomly selects a subgraph from the k-hop neighborhood of the target vertices. 
A mini-batch with two target vertices is shown in Figure~\ref{fig:comparison}(a). 
In the loading phase, the input features of the vertices in the bottom layer of the mini-batch are loaded into the GPUs.
During forward propagation, each GNN layer $l > 0$ aggregates and transforms the features of the vertices in the layer $l-1$ of the sample and produces the features of the vertices in the layer $l$ (see Figure~\ref{fig:minibatch}).
The last GNN layer computes the features of the target vertices, which are then used to calculate the loss. 
During backward propagation, the layers are executed in reverse order to compute gradients. 
Finally, all GPUs aggregate and apply the computed gradients.

\textbf{Redundant loading and computation}
Data parallelism is the most commonly used training strategy for mini-batch GNN training. 
In data parallel training, the target vertices are partitioned among GPUs, where each partition corresponds to a separate \textit{micro-batch} (see Figure~\ref{fig:comparison}(a)). 
Each GPU independently samples the neighborhood of the target vertices, loads the input features of all the vertices in the bottom layer of its micro-batch, and computes the hidden representation for the sampled nodes.
This approach has a high degree of sampling, data loading, and computational redundancy.
For example, in Figure~\ref{fig:comparison}(a), vertex $c$ is sampled in two micro-batches, the features of its neighbors are loaded by two GPUs, and its hidden representation is computed redundantly.

Table~\ref{tab:redundancy} further reports the degree of computational and data loading redundancy in data-parallel training.
With 4 GPUs, data parallelism creates 4 separate micro-batches (``Micro''), causing up to $1.2\times$ compute and $2.5\times$ feature loading compared to having only a single mini-batch (``Mini'').


\begin{table}[ht]
\centering
\small 
\tabcolsep=0.1cm
\begin{tabular}{|c|ccc|ccc|}
\hline
\multirow{2}{*}{\textbf{Graph}} & \multicolumn{3}{c|}{\textbf{\# Edges Computed}}                                                 & \multicolumn{3}{c|}{\textbf{\# Features Loaded}}                                                \\ \cline{2-7} 
                                & \multicolumn{1}{c|}{\textbf{ Micro}} & \multicolumn{1}{c|}{\textbf{ Mini}} & \textbf{Ratio} & \multicolumn{1}{c|}{\textbf{ Micro}} & \multicolumn{1}{c|}{\textbf{ Mini}} & \textbf{Ratio} \\ \hline
\textbf{Orkut}                  & \multicolumn{1}{c|}{926M}              & \multicolumn{1}{c|}{751M}             & 1.2x           & \multicolumn{1}{c|}{422M}              & \multicolumn{1}{c|}{169M}             & 2.5x          \\ \hline
\textbf{Papers100M}             & \multicolumn{1}{c|}{452M}              & \multicolumn{1}{c|}{389M}             & 1.2x          & \multicolumn{1}{c|}{231M}              & \multicolumn{1}{c|}{154M}             & 1.5x           \\ \hline
\textbf{Friendster}             & \multicolumn{1}{c|}{13.4B}             & \multicolumn{1}{c|}{13.1B}            & 1.0x          & \multicolumn{1}{c|}{11.4B}             & \multicolumn{1}{c|}{9.4B}             & 1.2x           \\ \hline
\end{tabular}%
\caption{Redundant computation and data loading. The total number of edges computed and feature data loaded over one epoch when each mini-batch is sampled as 4 micro-batches of size 1024 (Micro) vs. 1 mini-batch of size 4096 (Mini).}
\label{tab:redundancy}
\end{table}

\begin{SCfigure}
        \centering
            \caption{
            Breakdown of epoch time into sampling, loading, and training time for DGL, P3*, Quiver, and Edge (naive implementation of \tname) on Orkut with the GAT model. }
            \includegraphics[scale = .20]{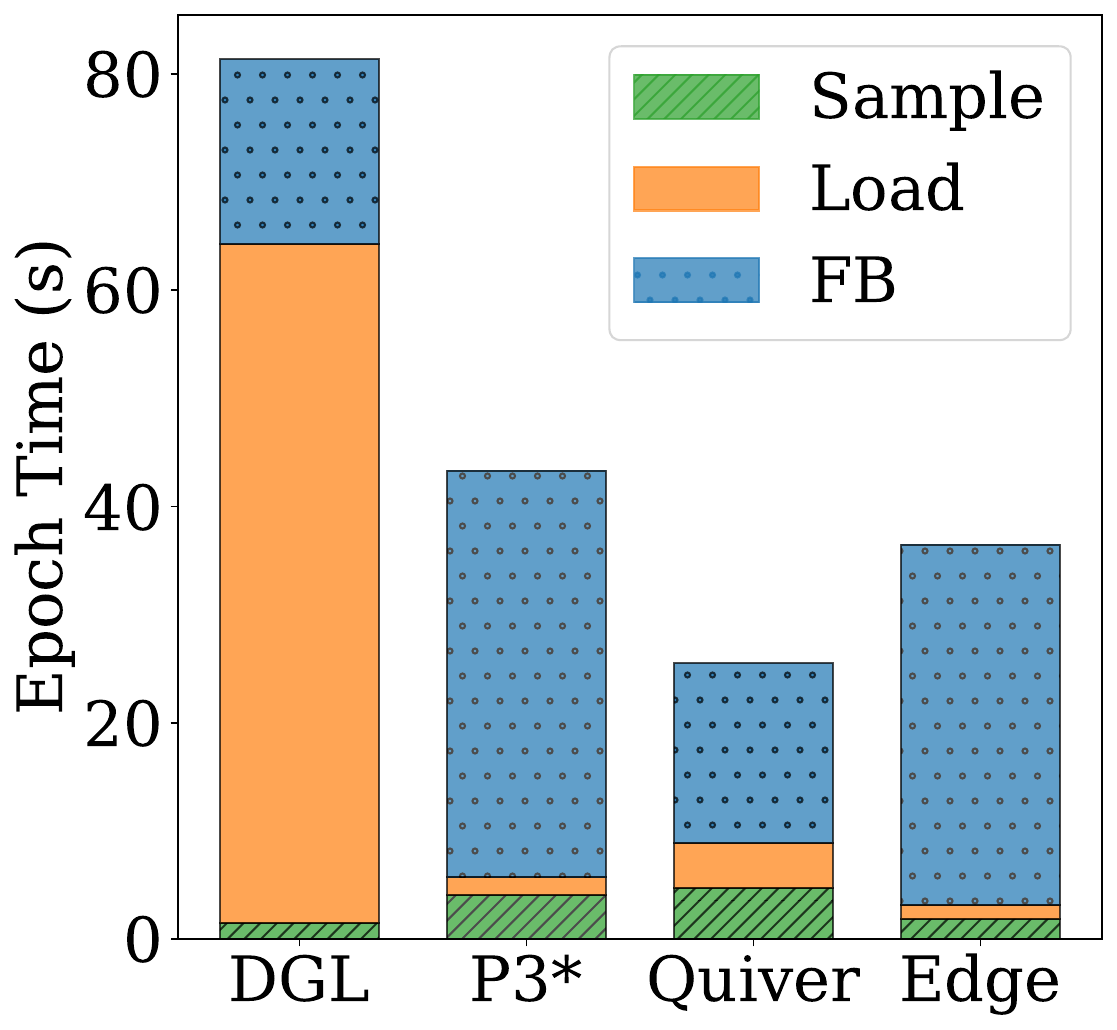}
\label{fig:orkut_epoch_breakdown}
\vspace{-10pt}
\end{SCfigure}

\mypar{Limitations of Existing Optimizations} 
The P3 system introduced a hybrid parallelism approach called \emph{push-pull parallelism} to partially address the redundancy problem of data parallelism~\cite{gandhi2021p3}. 
It avoids transferring input features among hosts by proposing an alternative to data parallelism called \emph{push-pull parallelism}.
The feature data for each vertex is partitioned across multiple GPUs.
As shown in Figure \ref{fig:comparison}(b), each GPU computes the first layer for \emph{all} the micro-batches on its local feature slice.
Then each GPU exchanges partial activations and continues the iteration for the remaining layers in a data-parallel fashion.
Figure \ref{fig:orkut_epoch_breakdown} shows that push-pull parallelism (described in Section \ref{sec:eval}) reduces data loading cost, but it also introduces an expensive shuffle that increases the overall training time.
This shuffle is expensive because $P3*$ still uses data parallelism for the upper layers.

Previous work also explored using GPU caching to reduce data-loading overhead in the context of data parallel training~\citep{pagraph, gnnlab, quiver, wholegraph, kalercommunicationeffgnn}.
These systems populate a static cache offline with frequently accessed input features.
Some systems use a distributed shared memory to enable GPUs to fetch features from other GPUs' memory using fast GPU-to-GPU interconnects like NVLink \citep{quiver, dsp, ugache}.
As shown in Figure~\ref{fig:orkut_epoch_breakdown}, the distributed shared-memory caching mechanism in Quiver~\citep{quiver} can reduce loading time for the Orkut graph, especially since its whose feature data can be fully cached across multiple GPUs. 

\sandremoved{
Distributed caching, however, does not fully address the performance issue of data loading, which can still take a significant fraction of the epoch time, as shown in Figure~\ref{fig:quiver_epoch_breakdown}.
For the GraphSage model, data loading over NVLink can be relatively expensive for Quiver with a graph that can be fully cached, like Orkut.
For larger graphs such as Papers100M, only a part of the input features can be cached on GPUs. 
Data loading can still stress the PCIe bus between the host and devices, resulting in unsatisfactory training performance. 
With Papers100M, only up to 60\% of the input features can be cached and the data loading time remains high. }

\mypar{Challenges of avoiding redundancy}
Despite these improvements, none of the above approaches avoids the computational and loading redundancies of mini-batch GNN training. 
In this work, we propose using \tname to avoid these redundancies by splitting each mini-batch sample on the fly at each iteration.
Splitting maps work associated with a vertex to only one GPU and shuffles vertex features across GPUs (see Figure \ref{fig:comparison}(c)).

Splitting is on the critical path, so it is necessary to minimize its overhead.
Splitting must also minimize communication costs during shuffles and balance the load across GPUs.
A na\"{i}ve approach would be to run a min-edge-cut graph partitioning algorithm on each mini-batch we sample.
However, this would be too computationally expensive since splitting must be executed during the sampling step of each iteration, and it is hard to parallelize across multiple GPUs.
A more practical approach would be to use graph partitions computed offline using a min-edge-cut graph partitioner like Metis~\cite{karypis1997metis}.
We evaluate this approach in Figure~\ref{fig:orkut_epoch_breakdown} and call it Edge, and find that it is not sufficient to outperform data parallelism since the partitioning is done on the entire graph, not on the specific mini-batch that is being split.
Edge performs worse than Quiver as the overhead of shuffling during training and load imbalance exceeds the savings from eliminating redundancy. 
This observation motivates us to design an optimal split assignment  (Section \ref{sec:workload}) that addresses these challenges. 

\section{\name: Training pipeline} \label{sec:opportunities}

\begin{figure*}[ht]
    \centering
    \includegraphics[scale=0.88, width=0.88\textwidth]{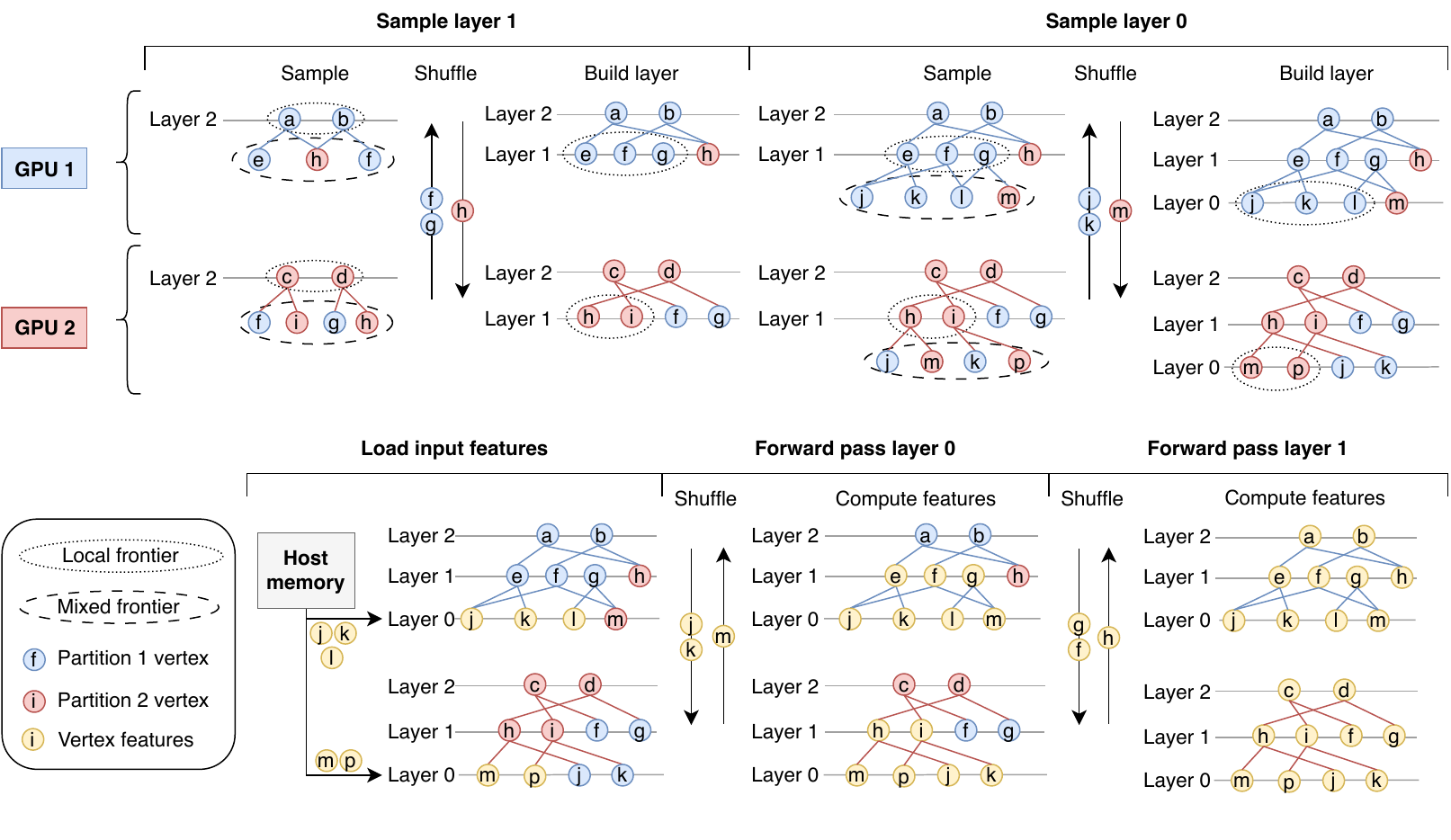}
    \vspace{-.5cm}
    \caption{Overview of the \name training pipeline.}
    \label{fig:overview}
\vspace{-.4cm}
\end{figure*}

To eliminate redundant work \emph{and} reduce training time, we introduce a hybrid parallelism approach called \textit{\tname}. 
Instead of sampling micro-batches with overlapping input and hidden vertices, one micro-batch for each GPU, the sampling phase now produces non-overlapping sets of vertices called \textit{splits}, each assigned to a different GPU.
The key to effective \tname is that the splits are obtained by a lightweight online \emph{splitting algorithm}, which provides probabilistic performance guarantees described in Section~\ref{sec:workload}, while sampling each mini-batch at each iteration.
Each GPU loads only the features of the input vertices within its assigned splits, taking into account each GPU's local cache.
During training, only one GPU computes the hidden features of any given vertex and then shuffles the computed feature to other GPUs.

\textbf{Example of a \name iteration}
We present a running example to describe one training iteration in \name (see Figure~\ref{fig:overview}) and introduce \name's API, such as the local and mixed frontier and the split index that supports existing efficient single-GPU kernels for sampling and training.
Although \name can be combined with distributed GPU caching, the example assumes no caching to simplify the description.

The first phase of each iteration is sampling.
\name pushes sampling to the GPU for performance reasons, in line with recent work~\citep{nextdoor,c-saw,wang2021skywalker,gsampler}.
Sampling proceeds layer-by-layer as each GPU samples its local split of the same mini-batch, rather than a separate micro-batch. 
At each layer, a GPU starts from a set of vertices called the {\em local frontier}, samples its neighbors, and obtains what we call the \emph{\ds}.
Unlike the local frontier, a \ds can include remote vertices.
For example, in Figure~\ref{fig:overview}, GPU~1 starts the iteration with the layer-2 local frontier $\{a,b\}$ and samples the \ds $\{e,h,f\}$, which includes the remote vertex $h$. 
Similarly, GPU 2 starts from the local frontier $\{c,d\}$ and samples the \ds $\{f,g,h,i\}$, which includes the two remote vertices $f$ and $g$.
Each GPU uses the splitting algorithm to separate the local and remote vertices and then shuffles the remote vertices to their partition.
GPUs then build the local frontier for the next layer based on the vertices they receive.
For example, the new local frontier of GPU 1 is $\{e,f,g\}$, where $\{g\}$ was sampled by GPU 2.
Analogously, the new local frontier of GPU 2 at layer-1 is $\{h,i\}$.
The next layer is the union of the new local frontier and the previous mixed frontier: $\{e, f, g, h\}$ for GPU 1 and $\{h, i, f, g\}$ for GPU 2.
Sampling for layer-1 is carried out in a similar way from its local frontier.
The splitting step creates an auxiliary data structure, called the \emph{shuffle index}, that consists of gather and scatter indexes to efficiently send and receive sparse vertex data during the shuffle rounds at each layer.

After sampling finishes, the loading step loads the input vertex features from the host memory into the GPUs.
When \tname is combined with caching, a GPU can skip loading an input feature if it is already cached locally.
Unlike data-parallel systems that could load redundant input features, \tname eliminates the redundant CPU-GPU data loading as splits do not have overlapping vertices. 
In the example of Figure~\ref{fig:overview}, we consider no caching, so GPU 1 loads the input features of the vertices in layer-0 local frontier $\{j,k,l\}$ from the host memory. Similarly, GPU 2 loads the input features of the vertices $\{m,p\}$.

The training phase begins after loading and proceeds bottom-up.
At each layer, each GPU is responsible for computing the hidden features of the vertices in its local frontier by aggregating the features of the neighbors in the layer below.
These neighbor vertices are the same vertices that were in the \ds for that layer during sampling, so we can reuse the shuffle index generated at that step.
In the example, GPU 1 must compute the layer-1 hidden features of vertices $\{e,f,g\}$.
To do that, it needs features of vertices $\{j,k,l,m\}$, which constitute the \ds 
 of layer-0 constructed during sampling and include the remote vertex~$h$.
GPU 1 sends the vertex features in its local frontier $\{j,k\}$ to GPU 2 and receives the features of the remote vertex $\{m\}$ by performing an all-to-all shuffle.
The shuffle rebuilds the \ds, which now holds all the hidden features required for the computation of all the vertices in the local frontier of layer-1 $\{e,f,g\}$.
GPU 1 proceeds to construct the mixed frontier of layer-1 by sending $\{g,f\}$ and receiving $\{h\}$ by performing an all-to-all shuffle. 
Finally, GPU-1 uses the mixed frontier $\{e,f,g,h\}$ of layer-1 to compute the hidden features of the target vertices $\{a,b\}$. 
The backward pass works in a similar way, moving from the top layer down to the bottom, but in the opposite direction. Importantly, the data follows the same paths as during the sampling step, so we can reuse the shuffle index.
\section{The Splitting Algorithm} \label{sec:workload}

Online splitting is challenging because it must be efficient, balance load across splits, and minimize communication cost, as discussed in Section~\ref{sec:problem}.
\name uses an embarrassingly parallel online splitting algorithm that maps each vertex to a split independently of each other in constant time. 
It does so by providing \emph{probabilistic} guarantees, rather than deterministic ones: given a random mini-batch sample, it minimizes the \emph{expected} communication cost while balancing the \emph{expected} load per split.
Formally, the splitting algorithm solves the following problem:

\noindent\textbf{Problem definition.} (Mini-batch splitting problem)
\emph{
Let $M(V_M,E_M)$ be a sampled mini-batch of a graph $G(V_G,E_G)$, $S$ a set of splits, and $f_M:\: V_M \rightarrow S$ a splitting function that assigns each vertex in $V_M$ to a split.
Let $S_i$ be the set of vertices assigned to split $i$ by $f_M$, $X_i$ a random variable expressing the number of vertices at layer $l > 0$ in $S_i$, $Y$ a random variable expressing the number of edges in $E_M$ having endpoints in two different splits, and $\epsilon \geq 0$ a tunable constant. The mini-batch splitting problem involves finding the splitting function $f_M$ that solves the following minimization problem:}
\begin{equation}
\begin{aligned}
\min_{f_M }\quad & \E[Y]\\
\textrm{s.t.} \quad & \forall i:\: \E[X_i] \leq (1 + \epsilon) \cdot \sum_{i \in [1,|S|]} \E[X_i] /|S|.
\end{aligned}
\vspace{-2pt}
\label{eqn:prob}
\end{equation}
The problem is to find a \emph{splitting function} $f_M$ that can be used online by the splitting algorithm to map each sampled vertex to a split, each corresponding to a different GPU device.
The random variables $X_i$ and $Y$ represent the computation and communication cost of the splits, respectively.
When the algorithm assigns a vertex at layer $l>0$ to a split, the corresponding GPU must sample its neighbors during the sampling phase and compute its hidden feature during the training phase.
Edges connecting vertices in different splits induce communication costs during the shuffle phases of both sampling and training. 
This problem is NP-hard since it can be reduced to min-edge-cut graph partitioning by selecting an appropriate sampling function.

\textbf{Splitting algorithm}
To avoid solving this NP-hard problem online, our approach is to reduce it to a problem that can be solved offline.
Therefore, we propose using a splitting algorithm that has an offline and an online part.
Offline, the algorithm finds a \emph{global partitioning function} $f_G:\: V_G \rightarrow D$ that statically maps each vertex in the whole input graph to a GPU device.
Then, online, the splitting algorithm uses $f_G$ as a substitute for $f_M$ to map each vertex to a device and thus to its corresponding split at each iteration.
Online splitting is embarrassingly parallel since $f_G$ maps each vertex to a split independently and does it in constant time.
\name uses the global partitioning $f_G$ also to determine the GPU where it statically caches input features of a vertex.
This ensures that the caches are consistent with the splits.

We now describe the details of the offline part of the splitting algorithm, which finds the global partitioning function.
The first stage of the offline algorithm is pre-sampling, which weighs the vertices and edges of the input graph.
Weights represent the computational and communication costs incurred by GPUs during \tadj sampling and training.
The second stage uses a weighted min-edge-cut graph partitioning algorithm to find the global partitioning function.

Given an input graph $G$, the pre-sampling stage assigns weights to the vertices and edges of $G$ to obtain a weighted graph $G_w$.
It runs the same sampling algorithm used during training for a fixed number of epochs.
At each iteration, the algorithm samples the k-hop neighborhood of the training (target) vertices in the mini-batch. 
It then assigns to each vertex $v$ a weight $w_V(v) = k_v/N$, where $k_v$ is the total number of times $v$ is sampled at a layer $l > 0$ across all samples and $N$ is the number of samples.
The weight of each edge $e$ is $w_E(e) = k_e/N$, where $k_e$ is the total number of times $e$ is sampled across all samples.

After completing the pre-sampling stage and obtaining the weighted graph $G_w$, the offline algorithm runs a weighted min-edge-cut graph partitioning algorithm on $G_w$ to obtain partitions.
The algorithm outputs partitions that minimize the sum of the weights of the edges in the cut and ensure that the load per partition, which is the sum of the weights of its vertices, is balanced. 
Formally, given a graph $G_w(V,E,w_V,w_E)$ and the number of partitions $d = |D|$, where $D$ is the set of GPU devices, a weighted min-edge-cut graph partition algorithm outputs the set of partitions $P=\{P_1,\ldots,P_d\}$ of $V$ that solves the following problem:
\begin{equation}
\begin{aligned}
\min_P \quad & \sum_{e \in C} w_E(e)\\
\textrm{s.t.} \quad & \forall i:\: L_i \leq (1 + \epsilon) \cdot L /d
\end{aligned}
\label{eqn:optim}
\end{equation}
where $w_E(e) = k_e/N$ is the edge weight of $e$, $C = \{\langle u, v \rangle \in E: u \in P_i, v \in P_j, i \neq j\}$ is the edge cut set, $L_i = \sum_{v \in P_i} w_V(v) = \sum_{v \in P_i} k_v/N$ is the load of each partition $P_i$, $L = \sum_{v \in V} w_V(v)$ is the total load across all partitions, and $\epsilon \geq 0$ is a tunable constant.
Note that since the minimization problem of Eq.~\ref{eqn:optim} is NP-hard, we use heuristics to solve it in practice, for example, using Metis~\citep{karypis1997metis}.
The partitions $P$ found by the graph partitioning algorithm determine the global partitioning function $f_G$ from vertices to GPUs.

\textbf{Analysis}
We show that our splitting algorithm finds a solution to the mini-batch splitting problem of Eq.~\ref{eqn:prob} by reducing it to the optimization problem of Eq.~\ref{eqn:optim} with $d=|S|$, which we can solve with a heuristic.
We show that $\E[X_i] = L_i$ and $\E[Y] = \sum_{e \in C} k_e/N$.
After doing that, we can conclude that Eq.~\ref{eqn:optim} minimizes $\E[Y]$ and constraints $\E[X_i]$ as in Eq.~\ref{eqn:prob}.

We start by showing that $\E[X_i] = L_i$.
Given a vertex $v \in V$, let the random variable $Z_v$ be the number of layers $l > 0$ where the vertex appears in the sample $S$. 
We have that $X_i = \sum_{v \in P_i} Z_v$, which implies that $\E[X_i]= \sum_{v \in P_i} \E[{Z_v}]$.
If we assign vertex weights using a sufficiently large number of samples $N$, according to the law of large numbers, we have $\E[{Z_v}] = k_v/N$, which implies that $\E[X_i] = L_i$.

We use a similar argument to show that $\E[Y] = \sum_{e \in C} k_e/N$.
Given an edge $e \in E$, let the random variable $Z_e$ be the number of times $e$ is sampled (at different layers) in $S$. 
The set of cross-split edges in the sample $S$ is a subset of the cross-partition edges $C$ in the whole graph, so $Y = \sum_{e \in C} Z_e$.
The expected value of $Y$ is given by $\E[Y] = \sum_{e \in C} \E[Z_e]$.
If we use a sufficiently large number of samples $N$ to assign edge weights, according to the law of large numbers, we have that $\E[Z_e] = k_e/N$, so $\E[Y] = \sum_{e \in C} k_e/N$.
\section{\name's API} \label{sec:api}
\name offers a \emph{layer-centric} programming API that simplifies the development of GNN sampling and training code and kernels by hiding cross-GPU coordination.
Besides simplicity, this approach supports the reuse of optimized single-GPU kernels for GNN sampling and training proposed by recent research~\citep{adaptgear, fastkernel, wu2021seastar, nextdoor, gsampler, fu2022tlpgnn, ye2023sparsetir, tcgnn}. 

\textbf{Motivation for a layer-centric API}
Single-GPU kernels used by data-parallel systems are layer-centric: they assume that at each GNN layer, the source vertex, destination vertex, and edge features for all incoming edges incident to the same destination vertex are locally available.
\name's API supports this property even after it splits destination vertices of the same mini-batch among different GPUs.
The alternative hybrid mini-batch parallelism approach, which is $P^3$, breaks this assumption to offer an \emph{edge-centric} API to enable pushing part of the computation of each micro-batch to multiple hosts and GPUs in a fine-grained manner.
For example, implementing state-of-the-art models such as Graph Attention Networks (GAT) requires a custom implementation to ensure correctness~\citep{gandhi2021p3}.

\textbf{Sampling} \label{sec:sampling}
Algorithm~\ref{algo:sp-sample} shows the pseudocode of sampling using \name's API. 

The \texttt{sp\_sample} function executed by each GPU takes as input the subset of target vertices in the mini-batch that are local to the GPU according to the splitting algorithm (\texttt{local\_targets}) and outputs a \emph{split}, which consists of the set of \texttt{edges} at each layer that are used by the local training kernels to aggregate features.
It also outputs a \textit{shuffle index} (\texttt{shuffle\_idx}), which consists of indices to efficiently send and receive vertex data in the frontiers during the sparse all-to-all shuffling at each layer.

The \texttt{sample\_layer} function outputs a new mixed frontier (\texttt{mixed\_front}) and the edges from the input local frontier to the output mixed frontier (\texttt{edges[l]}).
The mixed frontier includes both local and remote vertices.
The sampling code must invoke the \texttt{split} function to perform a shuffle and obtain a local frontier (Line~\ref{ln:split}).

The benefit of using this layer-wise sampling API is that we can directly use the sampling implementations in DGL~\citep{dgl} as well as other GPU sampling kernels~\citep{nextdoor,c-saw,wang2021skywalker,gsampler} that sample graph layer by layer.
\newcommand{\forcond}{\em l \KwTo [L-1, 0]}
\AlgoDontDisplayBlockMarkers\SetAlgoNoEnd\SetAlgoNoLine%
\begin{algorithm}[t]
    \caption{Split-parallel sampling.}
    \label{algo:sp-sample}
    \let\textnormal\ttfamily
    \footnotesize
    \DontPrintSemicolon
    \SetStartEndCondition{ }{}{}%
    \SetKwProg{Fn}{def}{\string:}{}
    \SetKwFunction{Range}{range}
    \SetKw{KwTo}{in}\SetKwFor{For}{for}{\string:}{}
    \SetKwFunction{dpsample}{dp\_sample}
    \SetKwFunction{splitsample}{sp\_sample}
    \SetKwFunction{sneigh}{\textbf{\color{teal}sample\_layer}}
    \SetKwFunction{unique}{unique}
    \SetKwFunction{union}{union}
    \SetKwFunction{local}{local\_unique}
    \SetKwFunction{concat}{concat}
    \SetKwFunction{csplit}{\textbf{\color{purple} add}}
    \SetKwFunction{split}{\textbf{\color{purple} split}}
    \SetKwFunction{localfrontier}{local\_frontier}
    \SetKwFunction{dst}{dst}
    \SetKwFunction{src}{src}
    \SetKwFunction{add}{add}
    \SetKwFunction{remap}{remap}    
    \SetKwFunction{getall}{getAll}    
    \SetKwFunction{bsplit}{\textbf{create\_sp\_frontier}}
    \SetKwFunction{shuffleidx}{shuffle\_idx}
    \Fn{\splitsample{\em local\_targets}}{
        local\_front[L] = local\_targets\; \\
        \For{\forcond}{
            mixed\_front, edges[l] = \sneigh{local\_front[l+1]}\label{ln:sample}\;
            local\_front[l], edges[l], shuffle\_idx[l] = \split{mixed\_front, edges[l]}\label{ln:split}\;
        }
        \Return edges, \shuffleidx\;
    }    
\end{algorithm}

\begin{algorithm}[t]
    \caption{Forward pass of split-parallel training.}
    \label{algo:sp-forward}
    \let\textnormal\ttfamily
    \footnotesize
    \DontPrintSemicolon
    \SetStartEndCondition{ }{}{}%
    \SetKwProg{Fn}{def}{\string:}{}
    \SetKwFunction{Range}{range}
    \SetKw{KwTo}{in}\SetKwFor{For}{for}{\string:}{}
    \SetKwFunction{dpfwpass}{dp\_forward}
    \SetKwFunction{splitfwpass}{sp\_forward}
    \SetKwFunction{local}{local\_unique}
    \SetKwFunction{csplit}{create\_upper\_split}
    \SetKwFunction{split}{{\color{turquoise} split}}
    \SetKwFunction{dst}{dst}
    \SetKwFunction{src}{src}
    \SetKwFunction{add}{add}
    \SetKwFunction{remap}{remap}    
    \SetKwFunction{getall}{getAll}    
    \SetKwFunction{shuffle}{\textbf{\color{purple}shuffle}}
    \SetKwFunction{splitidx}{shuffle\_idx}

    \Fn{\splitfwpass{\em model, feat, edges, \splitidx}}{
        hidden = feat\; \\ \label{ln:input}
        \For{\forcond}{
            \textbf{\color{teal} gnn\_layer} = model.layer(l)\;
            mixed\_hidden = \shuffle{\splitidx[l], local\_hidden} \label{ln:shuffle}\;    \\        local\_hidden = \textbf{\color{teal} gnn\_layer}(edges[l], mixed\_hidden)\label{ln:layer}\;
        }
        \Return hidden;
    }
    \vspace{-1pt}
\end{algorithm}

\textbf{Training}
\label{sec:training}
We now explain how to implement cooperative split-parallel training.
Algorithm~\ref{algo:sp-forward} shows the pseudocode for the forward propagation. 
The backward propagation works similarly, except that the computation happens from the top layer to the bottom layer. 

The \texttt{sp\_forward} function takes as input the GNN model (\texttt{model}) the input features of the vertices in the local split (\texttt{feat}), the structure of the split (\texttt{edges}), and the shuffle index (\texttt{shuffle\_idx}) required for each layer.
The latter two inputs are produced by the sampling code of Algorithm~\ref{algo:sp-sample}.
It outputs the hidden features of the target vertices.

At each layer $l$, each GPU starts by shuffling the features/activations of its local vertices to other GPUs using the \texttt{shuffle} function (Line~\ref{ln:shuffle}). 
The output of this function is the \texttt{mixed\_hidden} tensor, which contains all the features required to compute the hidden features at layer $l+1$, including the features of remote vertices.
The GNN layer then computes the next-layer hidden features for the vertices in the local partition of the GPU (Line~\ref{ln:layer}).

During the backward pass, each GPU computes the gradients for both local and remote vertices at each layer \texttt{l}. 
The shuffle index (\texttt{shuffle\_idx}) is then used to push gradients of remote vertices back to the GPU, storing the vertices in its local partition. 
The backward pass proceeds for each layer, reversing the direction of communication in the forward pass. 
Our abstraction thus ensures that gradients are computed accurately across layers when using \tname between GPUs and is agnostic to the specific GNN model being used. 

Similar to the case of sampling, using a layer-wise API for training allows us to reuse the optimized single-GPU training kernels ~\citep{adaptgear, fastkernel, wu2021seastar, fu2022tlpgnn, ye2023sparsetir} in \texttt{gnn\_layer}.
\section{Evaluation}
\label{sec:eval}
In this section, we evaluate \name by answering the following questions: 
What are the end-to-end speedups that can be achieved by \name relative to the baselines (§~\ref{sec:speedups})?
What is the impact of using \name's splitting algorithm, which provides probabilistic performance guarantees (§~\ref{sec:eval-split})?
How does \name scale to a larger number of GPUs within one host and across hosts (§~\ref{sec:eval-scalability})? 
How does performance vary when we vary the hyperparameters (\S~\ref{sec:ablation})?
How does \name's \tname impact accuracy (\S~\ref{sec:accuracy})?

\subsection{Experiment Settings}
\label{sec:settings}

\textbf{GNN models and datasets}
We consider two popular and diverse GNN models: GraphSage~\citep{graphsage} and GAT~\citep{velivckovic2017graph}.
We use the standard neighborhood sampling algorithm.
Its low computational complexity makes it less likely to hide the cost of shuffling during \tadj sampling.
By default, we use a sampling fanout of 15, 3 GNN layers, a default hidden size of 256 as used in \citep{graphsage}, and a batch size of 1024. 
\name's \tadj implementations use the same sampling and training kernels as DGL's data-parallel one.

We use three large datasets listed in Table~\ref{tab:dataset}.
The Paper100M dataset, derived from the directed citation graph, is the largest homogeneous dataset from the Open Graph Benchmark (OGB), a standard benchmark for GNN training ~\citep{ogb-node-dataset}.
Additionally, we utilize two other large synthetic graphs from the SNAP repository \citep{snapnets}, consisting of undirected social networks that are frequently used for evaluating GNN training performance.
 
\begin{table}[t]
    \centering
    \scalebox{.8}{
    \begin{tabular}{|l||c|c|c|c|} \hline
         \textbf{Dataset} & \textbf{\# Nodes} & \textbf{\# Edges} & \textbf{\# Feat} & \textbf{\# Type} \\ \hline \hline
         Orkut & 3.1M & 120M & 512 & Undirected\\ 
         Papers100M & 111M & 1.6B & 128 & Directed\\ 
         Friendster &  65M & 1.9B & 128 & Undirected\\ \hline
    \end{tabular}
    }
    \caption{Datasets used for the evaluation}
    \label{tab:dataset}
\end{table}

\textbf{Hardware setup} 
By default, our experiments use an AWS EC2 p3.8xlarge instance with 4 NVIDIA V100 GPUs (16GB memory) and Xeon E5-2686 v4  @ 2.70GHz,  with 32 CPU cores and 244 GB RAM. 
GPUs are connected to the CPU with a PCIe 3.0  16 bus and to each other via NVLink.
For experiments with 8 GPUs, we use a similar p3.16xlarge instance having 64 CPU cores and 488 GB of RAM.

\begin{table*}[]
\centering
\scalebox{.85}{
\begin{tabular}{|c|c|c|c|c|c|c|c|c|c|c|c|}
\hline 
\multirow{2}{*}{ Graph } & \multirow{2}{*}{ System } & \multicolumn{5}{c|}{\textbf{GraphSAGE}} & \multicolumn{5}{c|}{\textbf{GAT}} \\
\cline{3-12}
& & S & L & FB & Total(s) & Speedup & S & L & FB & Total(s) & Speedup \\ \hline
\multirow{5}{*}{ Orkut } & DGL & 1.5 & 62.7 & 9.2 & 73.4 & 4.4$\times$& 1.5 & 62.8 & 17.1 & 81.4 & 3.6$\times$\\
& P3* & 4.0 & 1.5 & 8.5 & 14.1 & 0.8$\times$& 4.0 & 1.7 & 37.6 & 43.3 & 1.9$\times$\\
& Quiver & 4.9 & 4.3 & 8.7 & 17.8 & 1.1$\times$& 4.7 & 4.2 & 16.4 & 25.5 & 1.1$\times$\\
& Edge & 1.9 & 1.3 & 25.1 & 28.3 & 1.7$\times$& 1.9 & 1.3 & 33.3 & 36.5 & 1.6$\times$\\ 
& \textbf{\name} & 1.9 & 0.1 & 14.8 & 16.7 & & 1.9 & 0.1 & 20.5 & 22.5 & \\  \hline
\multirow{5}{*}{ Papers100M } & DGL & 4.6 & 9.5 & 11.3 & 25.4 & 1.4$\times$& 4.7 & 9.0 & 31.7 & 45.4 & 1.2$\times$\\
& P3* & 3.3 & 11.5 & 25.8 & 40.6 & 2.2$\times$& 3.3 & 11.3 & 65.7 & 80.4 & 2.2$\times$\\
& Quiver & 11.8 & 10.4 & 11.7 & 34.7 & 1.9$\times$& 11.0 & 11.0 & 30.5 & 53.5 & 1.4$\times$\\
& Edge & 11.7 & 0.1 & 16.3 & 28.1 & 1.5$\times$& 11.5 & 0.1 & 40.1 & 51.7 & 1.4$\times$\\
& \textbf{\name} & 3.9 & 2.6 & 11.8 & 18.3 & & 3.8 & 2.3 & 31.1 & 37.2 & \\ \hline
\multirow{5}{*}{ Friendster } & DGL & 62.7 & 283.4 & 61.1 & 407.2 & 2.9$\times$& 62.6 & 284.8 & 245.9 & 593.3 & 1.7$\times$\\
& P3* & 85.9 & 350.8 & 151.5 & 588.1 & 4.1$\times$& 76.5 & 351.4 & 613.8 & 1041 & 3.0$\times$\\
& Quiver & 132.5 & 24.9 & 63.6 & 223.9 & 1.6$\times$& 135.5 & 24.8 & 243.5 & 404.2 & 1.2$\times$\\
& Edge & 65.7 & 1.0 & 121.8 & 188.5 & 1.3$\times$& 106.1 & 0.7 & 368.2 & 475.0 & 1.4$\times$\\
& \textbf{\name} & 41.2 & 2.2 & 98.9 & 142.3 & & 62.1 & 2.2 & 283.5 & 347.8 & \\ \hline
\end{tabular}
}
\caption{Epoch time (in seconds).  S = Sampling, L = Loading, FB = Forward and backward pass. The speedups are the total epoch time of other systems relative to \name. 
}
\vspace{-10pt}
\label{tab:main_nvlink}
\end{table*}

\textbf{Baseline systems}
We consider the following systems as baselines.
All systems perform synchronous training to avoid biasing model accuracy and use GPU-based sampling.
\begin{itemize}
\item \textbf{DGL} is a standard production library for data-parallel GNN training~\citep{dgl}.
We use DGL version 1.1.3, the same one we use as a component of \name. 
DGL only supports caching input features and the graph topology when they fully fit into one GPU.

\item \textbf{Quiver} is a recent data-parallel GNN training system that uses distributed caches and leverages fast direct GPU-GPU buses like NVLink~\citep{quiver}.
We use version 0.1.1.
Quiver supports distributed and partial caching across multiple GPUs.
\item \bm{$P^3$} is a distributed GNN training system that uses hybrid push-pull parallelism.
Its source code is not publicly available, so we adapt the push-pull parallelism approach to a single-host multi-GPU system and refer to our implementation as \textbf{P3*}. 

\item \textbf{Edge} is a variant of \name used to investigate the impact of using a na\"ive offline splitting algorithm that does not weigh vertices and edges using pre-sampling (see Section~\ref{sec:eval-split}). It uses min-cut partitioning and balances the number of edges and target vertices in each partition, as commonly done in data-parallel GNN training systems~\citep{distdgl}, while minimizing the number of edges across partitions.
\end{itemize}

\vspace{-10pt}
We configure all systems to maximize the memory available for caching the graph structure and input features while allocating sufficient memory to sample and train without going out of memory.
We configure Quiver
and \name to use the same sampling frequency criterion to rank the input features to cache as proposed in~\citep{gnnlab}.
P3* cannot cache input features for only a subset of the vertices, so it only uses caching for the Orkut graph.

\subsection{End-to-End Performance}
\label{sec:speedups}
We now compare the performance of \name with existing work: DGL, Quiver, and P3*. 
The comparison is based on epoch time only, as none of these systems biases the GNN model accuracy they train.
We measure the total epoch time and break it down for the three steps of the mini-batch training iterations: sampling a subgraph, which also includes splitting for \name, loading the input features into each GPU, and performing the forward and backward pass.

We report the results in Table~\ref{tab:main_nvlink}.
Overall, \name outperforms DGL by up to 4.4x (2.5x on average), P3* by up to 4.1x (2.4x on average), and Quiver by up to 1.9x (1.4x on average) by eliminating redundant loading and computation. 
By avoiding redundant computation, \name can reduce its sampling and training costs, mitigate the additional cost of shuffling, and in some cases be even faster than some data-parallel systems in those steps.

\textbf{Sampling time comparison}
The sampling step in \name entails not only sampling the mini-batch, as in the other systems, but also splitting the mini-batch, constructing the shuffle indexes, and shuffling vertices.
The evaluation shows that these additional costs are balanced, and sometimes offset, by the elimination of redundant work and by the use of distributed caching for the graph structure.
\name's online splitting is not a performance bottleneck because it is embarrassingly parallel and fast.

\textbf{Loading time comparison}
Compared to data parallel systems \name reduces the input feature loading time by reducing redundant loads.
P3* and Quiver perform much better than DGL in the Orkut graph because its input features can be fully cached across GPUs.
The input features for the Papers100M and Friendster graphs cannot be fully cached, even when using a distributed cache.
Quiver has lower loading times than DGL and P3*  for Friendster because it supports caching only a subset of the features in the GPUs, but it cannot leverage its cache effectively for Papers100M because of the high cost of loading feature cache misses from the host memory.

\textbf{Forward/backward pass time comparison}
In contrast to data-parallel systems, hybrid-parallel approaches such as P3* and \name introduce communication overhead in the form of shuffles during training, which generally leads to increased forward-backward (FB) times. When training the GraphSage model, P3* partially offsets the overhead of push-pull shuffling by distributing some computation of each micro-batch across all GPUs. This optimization allows P3* to achieve the shortest training time among all systems on the Orkut graph. However, in all other scenarios, P3* exhibits the longest FB times due to the need to shuffle all partial activations across micro-batches, a cost that becomes particularly pronounced with complex models like GAT, which generate larger intermediate activations.

\name addresses the shuffling overhead by eliminating redundant computation of hidden features, thereby improving overall efficiency. Nonetheless, its FB times remain higher than those of data-parallel systems. This gap is less pronounced for computationally intensive models such as GAT, where the benefits of avoiding redundant computation are more significant. Compared to P3*, \name consistently achieves lower training times across all cases except for the Orkut-GraphSage setting, attributed to its reduced shuffle costs and more efficient use of computation.

\subsection{Evaluation of the Splitting Algorithm}
\label{sec:eval-split}
\textbf{End-to-end impact of probabilistic splitting}
\name relies on an offline graph partitioning algorithm to provide probabilistic performance guarantees: balancing the expected load across splits and minimizing the expected communication costs across partitions.
We evaluate the impact of using this algorithm by combining \name's online splitting with three alternative offline partitioning algorithms that do not provide these guarantees. 

The \emph{\name} is the pre-sampling-based algorithm with probabilistic guarantees described in Section~\ref{sec:workload}.
The \emph{Node} algorithm partitions the graph using only the pre-sampled node weights. Comparing it to \emph{\name} shows the impact of using edge weights during graph partitioning.
The \emph{Edge} algorithm uses min-cut partitioning, but it does not assign weights to vertices and edges using pre-sampling. It balances the number of edges and target vertices in each partition, as commonly done in data-parallel GNN training systems~\citep{distdgl} while minimizing the number of edge cuts across partitions.
Finally, \emph{Rand} partitioning algorithm randomly assigns each vertex to a partition. 

Table~\ref{tab:main_nvlink} shows the end-to-end performance benefit of using \name's splitting algorithm compared to the Edge baseline.
\name helps improve the end-to-end training performance by up to 1.5$\times$ on Orkut, 1.7$\times$ on Papers100M, and 1.4$\times$ on Friendster.
\begin{figure}[ht]
    \centering
    \begin{subfigure}[t]{0.47\linewidth}
        \centering
        \includegraphics[width=\textwidth]{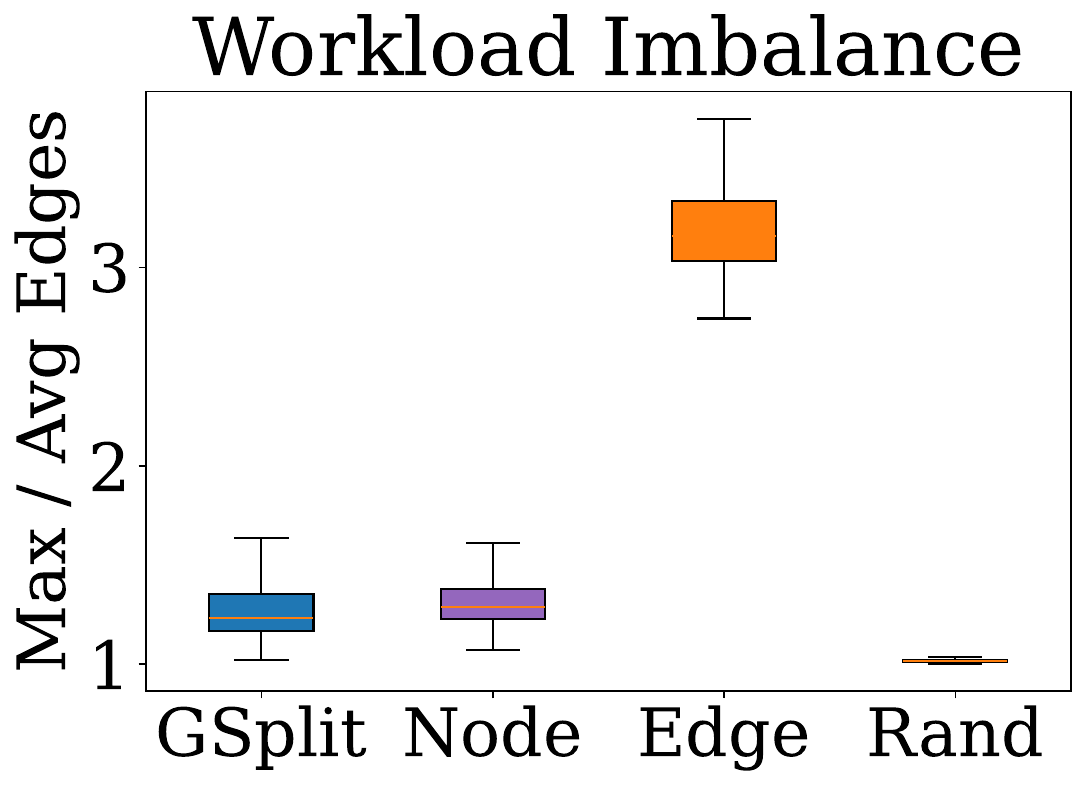}
        \label{fig:gsplit_partition_vs_others_workload}
    \end{subfigure}
    \hfill
    \begin{subfigure}[t]{0.5\linewidth}
        \centering
        \includegraphics[width=\textwidth]{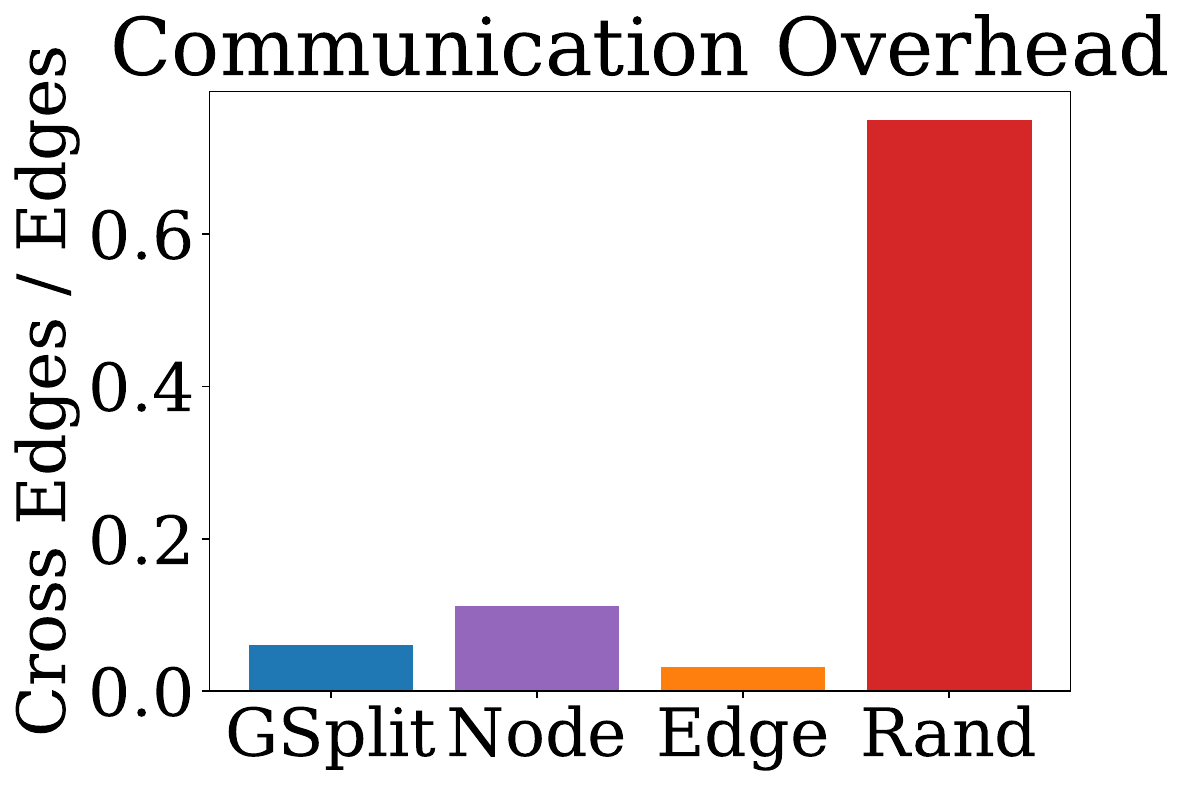}
        \label{fig:gsplit_partition_vs_others_comm}
    \end{subfigure}
    \vspace{-.6cm}
    \caption{\name vs. other offline partitioning algorithms.}
    \label{fig:gsplit_partition_vs_others}
\end{figure}
We analyze the reasons for these speedups more in-depth in Figure~\ref{fig:gsplit_partition_vs_others}, which compares the workload imbalance and communication costs of different partition strategies using the Papers100M graph. 
We quantify the \textit{workload imbalance} among the splits in each iteration as the maximum number of edges at layer $l >0 $ per split divided by the average, and the \textit{communication cost} as the percentage of cross-edges among splits over all the edges in the mini-batch.

As shown in Figure~\ref{fig:gsplit_partition_vs_others}, the \emph{Rand} baseline leads to the most evenly distributed computation cost across partitions. 
Yet, it results in a high communication overhead with 75\% of the edges crossing two partitions in most iterations. 
The \emph{Edge} baseline achieves a much lower edge cut and reduces the communication overhead. However, balancing the target vertices alone does not guarantee that the splits of the sampled mini-batches will be balanced.

The splitting algorithm of \name achieves the benefit of both approaches. 
It has a lower communication overhead compared to the random partition algorithm and a more balanced workload than simply balancing the number of target vertices in each partition. 
This is thanks to its offline pre-sampling approach.

In addition, we observe that assigning weights to edges using \name effectively reduces the communication overhead in a mini-batch.
Compared to Node, which does not weigh edges, the average ratio of cross edges over total edges is reduced from 9\% to 5\% for Papers100M as shown in Figure~\ref{fig:gsplit_partition_vs_others}. Better yet, the reduction in communication costs does not significantly impact workload imbalance. We observe a similar trend in Orkut and Friendster. 

\mypar{Offline pre-processing costs}
The splitting algorithm has two offline steps: pre-sampling and graph partitioning.
Empirically, we found that running ten epochs of sampling during the pre-sampling stage is sufficient.
Using a larger number of sampling epochs has little impact on load balancing and communication costs. 
When using $30$ and $100$ pre-sampling epochs, the difference in average load imbalance per mini-batch remains within 2\% for all the graphs, while the percentage of cross edges over the total number of edges per mini-batch remains within 7\% for Orkut, 2\% for Papers100M, and Friendster. 
Pre-sampling is fast relative to the overall training time.
The pre-sampling time is 19s for Orkut, 20s for Papers100M, and 288s for Friendster when using a machine with four RTX 3090 GPUs. 

The final offline step of \name is graph partitioning, which is commonly used in many distributed mini-batch GNN training systems.
We use METIS~\citep{mtmetis2013ipdps} to partition the graphs on an AWS r7a.x24large instance, which has 48 cores (96 threads) and 768GB of memory. The partitioning time is 14s for Orkut, 78s for Papers100M, and 534s for Friendster.
Both pre-sampling and partitioning are one-time costs that can be amortized by training over the same dataset multiple times.

\subsection{Scalablity}
\label{sec:eval-scalability}
We now evaluate the performance of \name against baseline systems across different configurations of GPUs and hosts. 

\begin{figure*}[t]
    \centering
    \includegraphics[width=\textwidth]{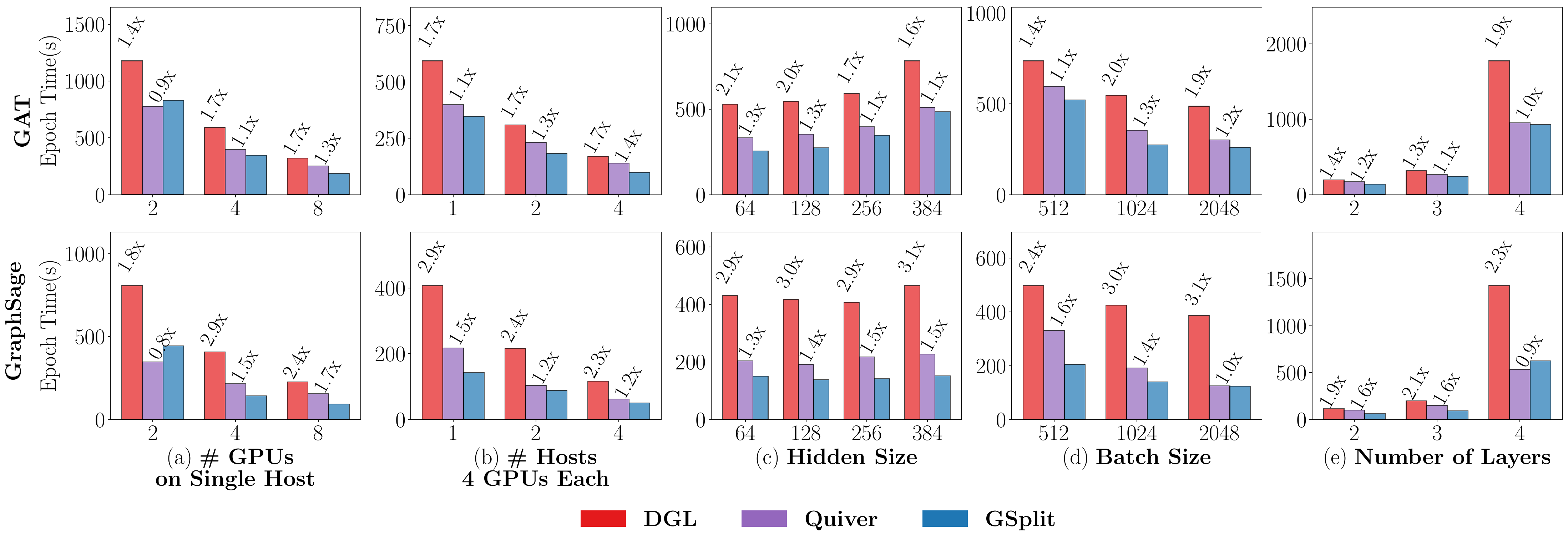}
   \vspace{-10pt} 
  \caption{Scalability and ablation study. The reported speedups are the epoch time of other systems relative to \name.} 
    \label{fig:ablation}
\end{figure*}

\textbf{Single-host.} 
We first evaluate using a single host and varying the number of GPUs in Figure~\ref{fig:ablation}(a). 
\name scales better than the other systems with a larger number of GPUs because it has more opportunities to avoid redundant loads and computation.
It can also make more efficient use of the GPU caches thanks to its use of collective communication primitives.
Quiver relies on direct remote memory access to transfer cached input features across GPUs efficiently.
This, however, is only possible between GPUs that have direct NVLink connections.
In our 8-GPU host, not all GPUs are directly connected with each other. Quiver circumvents this problem by replicating cached features across GPUs that have no direct links.
\name, by contrast, does not need to cache features redundantly.

\textbf{Multi-hosts.} We also run distributed multi-host experiments where each host has 4 GPUs and show the results in Figure~\ref{fig:ablation}(b).
\name can be implemented across multiple hosts by partitioning the graph and node structure across hosts.
Its shuffle function can operate across a diverse set of interconnects as it completely abstracts data collection into contiguous memory and utilizes low-level libraries such as NCCL for communication. 
As NCCL utilizes the best available interconnect, \name can transparently use high-speed interconnects such as the fifth-generation NVLink Switch technology.
However, in the absence of high-speed interconnects across hosts, as in our evaluation setting, \name uses a hybrid approach to scale to multiple hosts, using data parallelism across hosts and \tname within each host.
We observe that \name shows consistent speedups in all configurations and models.

\subsection{Ablation Study}
\label{sec:ablation}

We now estimate how consistent \name's speedups are when the model and training hyperparameters change.
The results for the Friendster graph are reported in Figure~\ref{fig:ablation}.

\textbf{Hidden size.} Increasing the hidden feature size impacts the FB time of \name negatively, increasing the overall volume of data shuffled. However, it also increases the gains of avoiding redundant computation, especially for complex models such as GAT. 
The two factors balance out and \name shows consistent speedups over the baselines, as shown in Figure~\ref{fig:ablation}(c).

\textbf{Batch size.}
We vary the mini-batch size while keeping a hidden size of 128 to avoid going out of memory. 
Larger mini-batches increase the relative cost of shuffling during the FB phase but also offer more opportunities to save on redundant data loading.
Overall, \name always outperforms the data-parallel baselines, as shown in Figure~\ref{fig:ablation}(d).

\textbf{Number of GNN Layers.}
In this experiment, we use a hidden size of 128 and pick the largest sampling fanout that avoids going out of memory for each number of layers.
The results are reported in Figure~\ref{fig:ablation}(e).
The GNNs are most commonly trained with 2 or 3 layers, i.e., 2 or 3 hops from the target vertices.
\name consistently outperforms the baselines in these settings.
Adding more layers in \tname increases the number of shuffles but also increases redundancy across minibatches. 

\subsection{Accuracy}
\label{sec:accuracy}

\begin{figure}[!ht]
    \centering
    \includegraphics[width=0.85\linewidth]{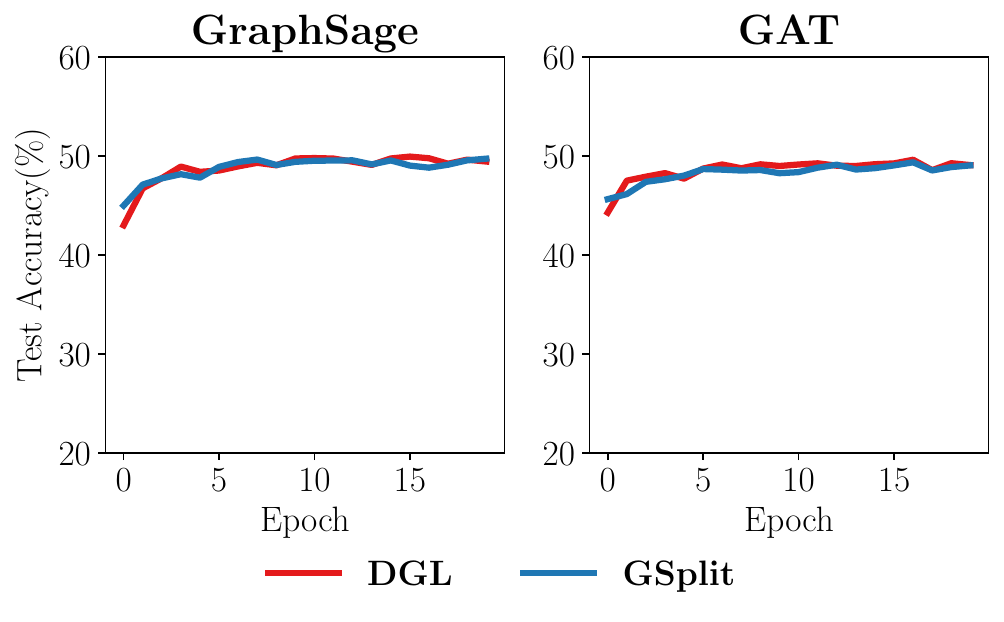}
    \vspace{-10pt} 
    \caption{Accuracy study}
    \label{fig:accuracy}
     \vspace{-10pt} 
\end{figure}

\name is a general GNN training system that introduces no system-level algorithmic bias that could affect training accuracy.
To validate this, we compare the test accuracy after each training epoch on the Papers100M dataset using the GraphSAGE and GAT models, trained with \name and the DGL data-parallel baseline.
Figure~\ref{fig:accuracy} shows that \name closely mirrors the baseline's accuracy across all epochs for both models, as expected.
These curves confirm the correctness of our split/shuffle abstraction, which automatically computes gradients across splits during the backward pass in a model-agnostic way.
\section{Related Work}

\textbf{Mini-batch training systems}
DistDGL~\citep{distdgl} and AliGraph~\citep{aligraph} use data-parallel mini-batch training to scale to large graphs.
ByteGNN optimizes distributed sampling for CPU-based data-parallel training~\citep{bytegnn}. 
Another research direction is taken by Marius++, which runs data-parallel GNN training on large graphs using a single GPU and an out-of-core approach rather than a distributed system~\citep{waleffe2022marius++}.
Pipelining multiple batches is another orthogonal technique to increase resource utilization and reduce training time~\citep{UVA-GNN-load,dsp,kaler2022acceleratingtraininginferencegraph}.

Prior work has focused on efficient single-GPU sampling with high-level programming APIs ~\citep{nextdoor,c-saw,wang2021skywalker,gsampler,tripathy2024distributedmatrixbasedsamplinggraph}.
GNNLab reduces the communication costs in distributed sampling using a factorized approach, where sampler GPUs cache the entire graph topology while other trainer GPUs only cache input features~\citep{gnnlab}.
Other systems propose caching both topological and feature data on each GPU~\citep{legion,ducati,dsp}.
DSP supports distributed sampling on a partitioned graph structure, but it performs two all-to-all shuffles for each sampled layer~\citep{dsp}.
\Tadj sampling requires only one all-to-all shuffle per sampled layer, since training is \tadj.
\Tname and cooperative training were introduced in a preliminary version of this paper~\citep{gsplit-arxiv}. Dedicated sampling algorithms can increase feature access locality~\citep{balin2023cooperative}.
APT adaptively chooses between data, push-pull, and split parallelism, but it does not consider probabilistic splitting~\cite{ma2025adaptive}.

\textbf{Sampling algorithms}
In our work, we consider general systems that support arbitrary sampling algorithms rather than imposing a specific algorithm, with its specific performance and accuracy tradeoffs, to the user.
A different line of work has focused on designing specific sampling algorithms that speed up GNN training and reduce data transfers at the cost of potentially biasing accuracy~\citep{dong2021global,twolevel, ramezani2020gcn,liu2023bgl, granndis, wan2022bns,song2023adgnn,zhang2023boostingdistributedfullgraphgnn}.

\textbf{Hybrid parallelism in full-graph training}
Workload characteristics of full graph training systems are different from data parallel minibatch training as the graph structure used in training does not change between iterations. 
Thus, overlapping optimized computation and communication schedules can be constructed during preprocessing as a fixed cost which is amortized during training ~\citep{roc, neugraph, cai2021dgcl, md2021distgnn, pipegcn, mgg, betty, G3, neutronstar, hongtu}.
Recent work has found mini-batch training to be generally more effective than full-graph training when using multiple GPUs~\cite{bajaj2024graphneuralnetworktraining}. 
\section{Conclusion}
This paper introduces \tname, a novel hybrid parallelism strategy for mini-batch training that mitigates redundant sampling, data loading, and computation commonly encountered in data parallelism and push-pull parallelism. 
A key technical challenge in using \tname is to develop a  lightweight splitting algorithms that can balance work and minimize communication. 
This paper proposes a probabilistic splitting algorithm that can push most of the computational overhead offline while achieving probabilistic guarantees.
It also shows that \tname can be applied transparently using a simple programming API.

\section*{Acknowledgements}
This material is based upon work supported by the National
Science Foundation under Grant No. CNS-2224054. Any opinions, findings, conclusions, or recommendations expressed in this material are those of the authors and do not necessarily reflect the views of the National Science Foundation. 
The work was also supported by Adobe Research grants and an Amazon Research Award. 
We would like to thank our anonymous reviewers and our shepherd Yuke Wang for their valuable feedback.

This manuscript has been authored by UT-Battelle, LLC, under contract DE-AC05-00OR22725 with the US Department of Energy (DOE). The US government retains and the publisher, by accepting the article for publication, acknowledges that the US government retains a nonexclusive, paid-up, irrevocable, worldwide license to publish or reproduce the published form of this manuscript, or allow others to do so, for US government purposes. DOE will provide public access to these results of federally sponsored research in accordance with the DOE Public Access Plan (https://www.energy.gov/doe-public-access-plan). This research used resources of the Oak Ridge Leadership Computing Facility at the Oak Ridge National Laboratory, which is supported by the Office of Science of the U.S. Department of Energy under Contract No. DE-AC05-00OR22725. 
\bibliography{ref}

@inproceedings{ma2025adaptive,
  title={Adaptive Parallel Training for Graph Neural Networks},
  author={Ma, Kaihao and Liu, Renjie and Yan, Xiao and Cai, Zhenkun and Song, Xiang and Wang, Minjie and Li, Yichao and Cheng, James},
  booktitle={Proceedings of the 30th ACM SIGPLAN Annual Symposium on Principles and Practice of Parallel Programming},
  pages={29--42},
  year={2025}
}

@inproceedings{ye2023sparsetir,
  title={SparseTIR: Composable abstractions for sparse compilation in deep learning},
  author={Ye, Zihao and Lai, Ruihang and Shao, Junru and Chen, Tianqi and Ceze, Luis},
  booktitle={Proceedings of the 28th ACM International Conference on Architectural Support for Programming Languages and Operating Systems, Volume 3},
  pages={660--678},
  year={2023}
}

@inproceedings{mtmetis2013ipdps,
  title={Multi-threaded graph partitioning},
  author={LaSalle, Dominique and Karypis, George},
  booktitle={Parallel \& Distributed Processing (IPDPS), 2013 IEEE 27th International Symposium on},
  pages={225--236},
  year={2013},
  organization={IEEE}
}

@inproceedings{fu2022tlpgnn,
  title={TLPGNN: A lightweight two-level parallelism paradigm for graph neural network computation on GPU},
  author={Fu, Qiang and Ji, Yuede and Huang, H Howie},
  booktitle={Proceedings of the 31st International Symposium on High-Performance Parallel and Distributed Computing},
  pages={122--134},
  year={2022}
}

@article{balin2023cooperative,
  title={Cooperative Minibatching in Graph Neural Networks},
  author={Balin, Muhammed Fatih and LaSalle, Dominique and {\c{C}}ataly{\"u}rek, {\"U}mit V},
  journal={arXiv preprint arXiv:2310.12403},
  month={Oct.},
  year={2023}
}

@article{gsplit-arxiv,
  title={{GSplit}: Scaling Graph Neural Network Training on Large Graphs via Split-Parallelism},
  author={Polisetty, Sandeep and Liu, Juelin and Falus, Kobi and Fung, Yi Ren and Lim, Seung-Hwan and Guan, Hui and Serafini, Marco},
  journal={arXiv preprint arXiv:2303.13775v1},
  month={Mar.},
  year={2023}
}

@inproceedings{liu2023bgl,
  title={{BGL}: {GPU-Efficient} {GNN} Training by Optimizing Graph Data {I/O} and Preprocessing},
  author={Liu, Tianfeng and Chen, Yangrui and Li, Dan and Wu, Chuan and Zhu, Yibo and He, Jun and Peng, Yanghua and Chen, Hongzheng and Chen, Hongzhi and Guo, Chuanxiong},
  booktitle={20th USENIX Symposium on Networked Systems Design and Implementation (NSDI 23)},
  pages={103--118},
  year={2023}
}

@article{song2023adgnn,
  title={ADGNN: Towards Scalable GNN Training with Aggregation-Difference Aware Sampling},
  author={Song, Zhen and Gu, Yu and Li, Tianyi and Sun, Qing and Zhang, Yanfeng and Jensen, Christian S and Yu, Ge},
  journal={Proceedings of the ACM on Management of Data},
  volume={1},
  number={4},
  pages={1--26},
  year={2023},
  publisher={ACM New York, NY, USA}
}

@article{wan2022bns,
  title={Bns-gcn: Efficient full-graph training of graph convolutional networks with partition-parallelism and random boundary node sampling},
  author={Wan, Cheng and Li, Youjie and Li, Ang and Kim, Nam Sung and Lin, Yingyan},
  journal={Proceedings of Machine Learning and Systems},
  volume={4},
  pages={673--693},
  year={2022}
}

@article{hongtu,
  title={HongTu: Scalable Full-Graph GNN Training on Multiple GPUs},
  author={Wang, Qiange and Chen, Yao and Wong, Weng-Fai and He, Bingsheng},
  journal={Proceedings of the ACM on Management of Data},
  volume={1},
  number={4},
  pages={1--27},
  year={2023},
  publisher={ACM New York, NY, USA}
}

@inproceedings{neutronstar,
author = {Wang, Qiange and Zhang, Yanfeng and Wang, Hao and Chen, Chaoyi and Zhang, Xiaodong and Yu, Ge},
title = {NeutronStar: Distributed GNN Training with Hybrid Dependency Management},
year = {2022},
isbn = {9781450392495},
publisher = {Association for Computing Machinery},
address = {New York, NY, USA},
url = {https://doi.org/10.1145/3514221.3526134},
doi = {10.1145/3514221.3526134},
booktitle = {Proceedings of the 2022 International Conference on Management of Data},
pages = {1301–1315},
numpages = {15},
location = {Philadelphia, PA, USA},
series = {SIGMOD '22}
}

@article{bytegnn,
author = {Zheng, Chenguang and Chen, Hongzhi and Cheng, Yuxuan and Song, Zhezheng and Wu, Yifan and Li, Changji and Cheng, James and Yang, Hao and Zhang, Shuai},
title = {ByteGNN: efficient graph neural network training at large scale},
year = {2022},
issue_date = {February 2022},
publisher = {VLDB Endowment},
volume = {15},
number = {6},
issn = {2150-8097},
url = {https://doi.org/10.14778/3514061.3514069},
doi = {10.14778/3514061.3514069},
journal = {Proc. VLDB Endow.},
month = {feb},
pages = {1228–1242},
numpages = {15}
}

@INPROCEEDINGS{fastkernel,
  author={Fan, Ruibo and Wang, Wei and Chu, Xiaowen},
  booktitle={2023 IEEE International Parallel and Distributed Processing Symposium (IPDPS)}, 
  title={Fast Sparse GPU Kernels for Accelerated Training of Graph Neural Networks}, 
  booktitle={2023 IEEE International Parallel and Distributed Processing Symposium (IPDPS)}, 
  year={2023}
}

@article{adaptgear,
  title={AdaptGear: Accelerating GNN Training via Adaptive Subgraph-Level Kernels on GPUs},
  author={Yangjie Zhou and Yaoxu Song and Jingwen Leng and Zihan Liu and Weihao Cui and Zhendong Zhang and Cong Guo and Quan Chen and Li Li and Minyi Guo},
  journal={Proceedings of the 20th ACM International Conference on Computing Frontiers},
  year={2023},
  url={https://api.semanticscholar.org/CorpusID:258960480}
}

@INPROCEEDINGS{twolevel,
  author={Zhang, Zhe and Luo, Ziyue and Wu, Chuan},
  booktitle={IEEE INFOCOM 2023 - IEEE Conference on Computer Communications}, 
  title={Two-level Graph Caching for Expediting Distributed GNN Training}, 
  year={2023},
  volume={},
  number={},
  pages={1-10},
  doi={10.1109/INFOCOM53939.2023.10228911}}

@article{G3,
author = {Wan, Xinchen and Xu, Kaiqiang and Liao, Xudong and Jin, Yilun and Chen, Kai and Jin, Xin},
title = {Scalable and Efficient Full-Graph GNN Training for Large Graphs},
year = {2023},
issue_date = {June 2023},
publisher = {Association for Computing Machinery},
address = {New York, NY, USA},
volume = {1},
number = {2},
url = {https://doi.org/10.1145/3589288},
doi = {10.1145/3589288},
journal = {Proc. ACM Manag. Data},
month = {jun},
articleno = {143},
numpages = {23}
}

@inproceedings{betty,
author = {Yang, Shuangyan and Zhang, Minjia and Dong, Wenqian and Li, Dong},
title = {Betty: Enabling Large-Scale GNN Training with Batch-Level Graph Partitioning},
year = {2023},
isbn = {9781450399166},
publisher = {Association for Computing Machinery},
address = {New York, NY, USA},
url = {https://doi.org/10.1145/3575693.3575725},
doi = {10.1145/3575693.3575725},
booktitle = {Proceedings of the 28th ACM International Conference on Architectural Support for Programming Languages and Operating Systems, Volume 2},
pages = {103–117},
numpages = {15},
location = {Vancouver, BC, Canada},
series = {ASPLOS 2023}
}

@inproceedings {mgg,
author = {Yuke Wang and Boyuan Feng and Zheng Wang and Tong Geng and Kevin Barker and Ang Li and Yufei Ding},
title = {{MGG}: Accelerating Graph Neural Networks with {Fine-Grained} {Intra-Kernel} {Communication-Computation} Pipelining on {Multi-GPU} Platforms},
booktitle = {17th USENIX Symposium on Operating Systems Design and Implementation (OSDI 23)},
year = {2023},
isbn = {978-1-939133-34-2},
address = {Boston, MA},
pages = {779--795},
url = {https://www.usenix.org/conference/osdi23/presentation/wang-yuke},
publisher = {USENIX Association},
month = jul
}

@inproceedings{pipegcn,
  title={{PipeGCN}: Efficient Full-Graph Training of Graph Convolutional Networks with Pipelined Feature Communication},
  author={Wan, C and Li, Y and Wolfe, Cameron R and Kyrillidis, A and Kim, Nam S and Lin, Y},
  booktitle={The Tenth International Conference on Learning Representations (ICLR 2022)},
  year={2022}
}

@inproceedings {legion,
author = {Jie Sun and Li Su and Zuocheng Shi and Wenting Shen and Zeke Wang and Lei Wang and Jie Zhang and Yong Li and Wenyuan Yu and Jingren Zhou and Fei Wu},
title = {Legion: Automatically Pushing the Envelope of {Multi-GPU} System for {Billion-Scale} {GNN} Training},
booktitle = {2023 USENIX Annual Technical Conference (USENIX ATC 23)},
year = {2023},
isbn = {978-1-939133-35-9},
address = {Boston, MA},
pages = {165--179},
url = {https://www.usenix.org/conference/atc23/presentation/sun},
publisher = {USENIX Association},
month = jul
}

@inproceedings{wholegraph,
author = {Yang, Dongxu and Liu, Junhong and Qi, Jiaxing and Lai, Junjie},
title = {WholeGraph: A Fast Graph Neural Network Training Framework with Multi-GPU Distributed Shared Memory Architecture},
year = {2022},
isbn = {9784665454445},
publisher = {IEEE Press},
booktitle = {Proceedings of the International Conference on High Performance Computing, Networking, Storage and Analysis},
articleno = {54},
numpages = {14},
location = {Dallas, Texas},
series = {SC '22}
}

@Inproceedings{gsampler,
 author = {Ping Gong and Renjie Liu and Zunyao Mao and Zhenkun Cai and Xiao Yan and Cheng Li and Minjie Wang and Zhuozhao Li},
 title = {gSampler: General and efficient GPU-based graph sampling for graph learning},
 year = {2023},
 booktitle = {ACM 2023 Symposium on Operating Systems Principles (SOSP)},
}

@article{ducati,
author = {Zhang, Xin and Shen, Yanyan and Shao, Yingxia and Chen, Lei},
title = {DUCATI: A Dual-Cache Training System for Graph Neural Networks on Giant Graphs with the GPU},
year = {2023},
issue_date = {June 2023},
publisher = {Association for Computing Machinery},
address = {New York, NY, USA},
volume = {1},
number = {2},
url = {https://doi.org/10.1145/3589311},
doi = {10.1145/3589311},
journal = {Proc. ACM Manag. Data},
month = {jun},
articleno = {166},
numpages = {24}
}

@inproceedings{ugache,
author = {Song, Xiaoniu and Zhang, Yiwen and Chen, Rong and Chen, Haibo},
title = {UGACHE: A Unified GPU Cache for Embedding-Based Deep Learning},
year = {2023},
isbn = {9798400702297},
publisher = {Association for Computing Machinery},
address = {New York, NY, USA},
url = {https://doi.org/10.1145/3600006.3613169},
doi = {10.1145/3600006.3613169},
booktitle = {Proceedings of the 29th Symposium on Operating Systems Principles},
pages = {627–641},
numpages = {15},
location = {Koblenz, Germany},
series = {SOSP '23}
}

@inproceedings{dsp,
author = {Cai, Zhenkun and Zhou, Qihui and Yan, Xiao and Zheng, Da and Song, Xiang and Zheng, Chenguang and Cheng, James and Karypis, George},
title = {DSP: Efficient GNN Training with Multiple GPUs},
year = {2023},
isbn = {9798400700156},
publisher = {Association for Computing Machinery},
address = {New York, NY, USA},
url = {https://doi.org/10.1145/3572848.3577528},
doi = {10.1145/3572848.3577528},
booktitle = {Proceedings of the 28th ACM SIGPLAN Annual Symposium on Principles and Practice of Parallel Programming},
pages = {392–404},
numpages = {13},
location = {Montreal, QC, Canada},
series = {PPoPP '23}
}

@inproceedings{wang2021skywalker,
  title={Skywalker: Efficient Alias-Method-Based Graph Sampling and Random Walk on GPUs},
  author={Wang, Pengyu and Li, Chao and Wang, Jing and Wang, Taolei and Zhang, Lu and Leng, Jingwen and Chen, Quan and Guo, Minyi},
  booktitle={2021 30th International Conference on Parallel Architectures and Compilation Techniques (PACT)},
  pages={304--317},
  year={2021},
  organization={IEEE}
}

@inproceedings{pagraph,
author = {Lin, Zhiqi and Li, Cheng and Miao, Youshan and Liu, Yunxin and Xu, Yinlong},
title = {PaGraph: Scaling GNN Training on Large Graphs via Computation-Aware Caching},
year = {2020},
isbn = {9781450381376},
publisher = {Association for Computing Machinery},
address = {New York, NY, USA},
doi = {10.1145/3419111.3421281},
booktitle = {Proceedings of the 11th ACM Symposium on Cloud Computing},
pages = {401–415},
numpages = {15},
location = {Virtual Event, USA},
series = {SoCC '20}
}

@inproceedings{dong2021global,
  title={Global Neighbor Sampling for Mixed CPU-GPU Training on Giant Graphs},
  author={Dong, Jialin and Zheng, Da and Yang, Lin F and Karypis, Geroge},
  booktitle = {Proceedings of the 27th ACM SIGKDD Conference on Knowledge Discovery \& Data Mining},
  pages = {289–299},
  numpages = {11},
  series = {KDD '21},
  year={2021}
}

@inproceedings{c-saw,
  title={C-SAW: a framework for graph sampling and random walk on GPUs},
author={Pandey, Santosh and Li, Lingda and Hoisie, Adolfy and Li, Xiaoye S. and Liu, Hang},
  booktitle={SC20: International Conference for High Performance Computing, Networking, Storage and Analysis}, 
  year={2020}
}

@article{aligraph,
  title={Aligraph: A comprehensive graph neural network platform},
  author={Zhu, Rong and Zhao, Kun and Yang, Hongxia and Lin, Wei and Zhou, Chang and Ai, Baole and Li, Yong and Zhou, Jingren},
  journal = {Proc. VLDB Endow.},
  volume = {12},
  number = {12},
  pages = {2094–2105},
  year={2019}
}

@inproceedings{nextdoor,
  title={Accelerating Graph Sampling for Graph Machine Learning using GPUs},
  author={Jangda, Abhinav and Polisetty, Sandeep and Guha, Arjun and Serafini, Marco},
  booktitle={European Conference on Computer Systems (EuroSys)},
  year={2021}
}

@article{roc,
  title={Improving the accuracy, scalability, and performance of graph neural networks with roc},
  author={Jia, Zhihao and Lin, Sina and Gao, Mingyu and Zaharia, Matei and Aiken, Alex},
  journal={Proceedings of Machine Learning and Systems},
  volume={2},
  pages={187--198},
  year={2020}
}

@article{distdgl,
  title={DistDGL: Distributed Graph Neural Network Training for Billion-Scale Graphs},
  author={Zheng, Da and Ma, Chao and Wang, Minjie and Zhou, Jinjing and Su, Qidong and Song, Xiang and Gan, Quan and Zhang, Zheng and Karypis, George},
  journal={arXiv preprint arXiv:2010.05337},
  year={2020}
}

@inproceedings{wu2021seastar,
  title={Seastar: vertex-centric programming for graph neural networks},
  author={Wu, Yidi and Ma, Kaihao and Cai, Zhenkun and Jin, Tatiana and Li, Boyang and Zheng, Chenguang and Cheng, James and Yu, Fan},
  booktitle={Proceedings of the Sixteenth European Conference on Computer Systems},
  pages={359--375},
  year={2021}
}

@inproceedings{velivckovic2017graph,
  title={Graph attention networks},
  author={Veli{\v{c}}kovi{\'c}, Petar and Cucurull, Guillem and Casanova, Arantxa and Romero, Adriana and Lio, Pietro and Bengio, Yoshua},
  booktitle={International Conference on Learning Representations},
  year={2018}
}

@misc{snapnets,
      author       = {Jure Leskovec and Andrej Krevl},
      title        = {{SNAP Datasets}: {Stanford} Large Network Dataset Collection},
      howpublished = {\url{http://snap.stanford.edu/data}},     
      year         = 2014
}

@article{pytorch-geometric,
  title={Fast graph representation learning with PyTorch Geometric},
  author={Fey, Matthias and Lenssen, Jan Eric},
  journal={arXiv preprint arXiv:1903.02428},
  year={2019}
}

@article{dgl,
      title={Deep Graph Library: A Graph-Centric, Highly-Performant Package for Graph Neural Networks}, 
      author={Minjie Wang and Da Zheng and Zihao Ye and Quan Gan and Mufei Li and Xiang Song and Jinjing Zhou and Chao Ma and Lingfan Yu and Yu Gai and Tianjun Xiao and Tong He and George Karypis and Jinyang Li and Zheng Zhang},
      year={2020},
      eprint={1909.01315},
      archivePrefix={arXiv},
      primaryClass={id='cs.LG' full_name='Machine Learning' is_active=True alt_name=None in_archive='cs' is_general=False description='Papers on all aspects of machine learning research (supervised, unsupervised, reinforcement learning, bandit problems, and so on) including also robustness, explanation, fairness, and methodology. cs.LG is also an appropriate primary category for applications of machine learning methods.'}
}

@inproceedings{graphsage,
author = {Hamilton, William L. and Ying, Rex and Leskovec, Jure},
title = {{Inductive Representation Learning on Large Graphs}},
year = {2017},
booktitle = {{Proceedings of the 31st International Conference on Neural Information Processing Systems}},
series = {NIPS’17}
}

@article{karypis1997metis,
  title={METIS: A software package for partitioning unstructured graphs, partitioning meshes, and computing fill-reducing orderings of sparse matrices},
  author={Karypis, George and Kumar, Vipin},
  year={1997}
}

@inproceedings{ramezani2020gcn,
  title={GCN meets GPU: Decoupling" When to Sample" from" How to Sample".},
  author={Ramezani, Morteza and Cong, Weilin and Mahdavi, Mehrdad and Sivasubramaniam, Anand and Kandemir, Mahmut T},
  booktitle={NeurIPS},
  year={2020}
}

@inproceedings{md2021distgnn,
  title={DistGNN: Scalable Distributed Training for Large-Scale Graph Neural Networks},
  author={Md, Vasimuddin and Misra, Sanchit and Ma, Guixiang and Mohanty, Ramanarayan and Georganas, Evangelos and Heinecke, Alexander and Kalamkar, Dhiraj and Ahmed, Nesreen K and Avancha, Sasikanth},
    booktitle = {SC'21: Proceedings of the International Conference for High Performance Computing, Networking, Storage and Analysis},
    year={2021}
}

@misc{ogb-node-dataset,
      title={Open Graph Benchmark: Datasets for Machine Learning on Graphs}, 
      author={Weihua Hu and Matthias Fey and Marinka Zitnik and Yuxiao Dong and Hongyu Ren and Bowen Liu and Michele Catasta and Jure Leskovec},
      year={2021},
      eprint={2005.00687},
      archivePrefix={arXiv},
      primaryClass={cs.LG}
}

@inproceedings{cai2021dgcl,
author = {Cai, Zhenkun and Yan, Xiao and Wu, Yidi and Ma, Kaihao and Cheng, James and Yu, Fan},
title = {DGCL: An Efficient Communication Library for Distributed GNN Training},
year = {2021},
isbn = {9781450383349},
publisher = {Association for Computing Machinery},
address = {New York, NY, USA},
doi = {10.1145/3447786.3456233},
booktitle = {Proceedings of the Sixteenth European Conference on Computer Systems},
pages = {130–144},
numpages = {15},
location = {Online Event, United Kingdom},
series = {EuroSys '21}
}

@inproceedings{gandhi2021p3,
  title={P3: Distributed Deep Graph Learning at Scale},
  author={Gandhi, Swapnil and Iyer, Anand Padmanabha},
  booktitle={15th {USENIX} Symposium on Operating Systems Design and Implementation ({OSDI} 21)},
  pages={551--568},
  year={2021}
}

@article{quiver,
  title={Quiver: Supporting gpus for low-latency, high-throughput gnn serving with workload awareness},
  author={Tan, Zeyuan and Yuan, Xiulong and He, Congjie and Sit, Man-Kit and Li, Guo and Liu, Xiaoze and Ai, Baole and Zeng, Kai and Pietzuch, Peter and Mai, Luo},
  journal={arXiv preprint arXiv:2305.10863},
  year={2023}
}

@article{waleffe2022marius++,
  title={Marius++: Large-Scale Training of Graph Neural Networks on a Single Machine},
  author={Waleffe, Roger and Mohoney, Jason and Rekatsinas, Theodoros and Venkataraman, Shivaram},
  journal={arXiv preprint arXiv:2202.02365},
  year={2022}
}

@inproceedings{neugraph,
  title={Neugraph: parallel deep neural network computation on large graphs},
  author={Ma, Lingxiao and Yang, Zhi and Miao, Youshan and Xue, Jilong and Wu, Ming and Zhou, Lidong and Dai, Yafei},
  booktitle={2019 {USENIX} Annual Technical Conference ({USENIX ATC} 19)},
  pages={443--458},
  year={2019}
}

@inproceedings{gnnlab,
  title={GNNLab: a factored system for sample-based GNN training over GPUs},
  author={Yang, Jianbang and Tang, Dahai and Song, Xiaoniu and Wang, Lei and Yin, Qiang and Chen, Rong and Yu, Wenyuan and Zhou, Jingren},
  booktitle={Proceedings of the Seventeenth European Conference on Computer Systems},
  pages={417--434},
  year={2022}
}

@article{UVA-GNN-load,
author = {Min, Seung Won and Wu, Kun and Huang, Sitao and Hidayeto\u{g}lu, Mert and Xiong, Jinjun and Ebrahimi, Eiman and Chen, Deming and Hwu, Wen-mei},
title = {Large Graph Convolutional Network Training with GPU-Oriented Data Communication Architecture},
year = {2021},
issue_date = {July 2021},
publisher = {VLDB Endowment},
volume = {14},
number = {11},
journal = {Proc. VLDB Endow.},
month = {oct},
pages = {2087–2100},
numpages = {14}
}

@inproceedings{tripathy2024distributedmatrixbasedsamplinggraph,
 author = {Tripathy, Alok and Yelick, Katherine and Bulu\c{c}, Ayd\i n},
 booktitle = {Proceedings of Machine Learning and Systems},
 editor = {P. Gibbons and G. Pekhimenko and C. De Sa},
 pages = {253--265},
 title = {Distributed Matrix-Based Sampling for Graph Neural Network Training},
 volume = {6},
 year = {2024}
}

@inproceedings{kaler2022acceleratingtraininginferencegraph,
 author = {Kaler, Tim and Stathas, Nickolas and Ouyang, Anne and Iliopoulos, Alexandros-Stavros and Schardl, Tao and Leiserson, Charles E. and Chen, Jie},
 booktitle = {Proceedings of Machine Learning and Systems},
 editor = {D. Marculescu and Y. Chi and C. Wu},
 pages = {172--189},
 title = {Accelerating Training and Inference of Graph Neural Networks with Fast Sampling and Pipelining},
 volume = {4},
 year = {2022}
}

@inproceedings{kalercommunicationeffgnn,
 author = {Kaler, Tim and Iliopoulos, Alexandros and Murzynowski, Philip and Schardl, Tao and Leiserson, Charles E. and Chen, Jie},
 booktitle = {Proceedings of Machine Learning and Systems},
 editor = {D. Song and M. Carbin and T. Chen},
 pages = {477--494},
 publisher = {Curan},
 title = {Communication-Efficient Graph Neural Networks with Probabilistic Neighborhood Expansion Analysis and Caching},
 volume = {5},
 year = {2023}
}

@misc{zhang2023boostingdistributedfullgraphgnn,
      title={Boosting Distributed Full-graph GNN Training with Asynchronous One-bit Communication}, 
      author={Meng Zhang and Qinghao Hu and Peng Sun and Yonggang Wen and Tianwei Zhang},
      year={2023},
      eprint={2303.01277},
      archivePrefix={arXiv},
      primaryClass={cs.DC},
      url={https://arxiv.org/abs/2303.01277}, 
}

@inproceedings{granndis,
author = {Song, Jaeyong and Jang, Hongsun and Lim, Hunseong and Jung, Jaewon and Kim, Youngsok and Lee, Jinho},
title = {GraNNDis: Fast Distributed Graph Neural Network Training Framework for Multi-Server Clusters},
year = {2024},
isbn = {9798400706318},
publisher = {Association for Computing Machinery},
address = {New York, NY, USA},
url = {https://doi.org/10.1145/3656019.3676892},
doi = {10.1145/3656019.3676892},
booktitle = {Proceedings of the 2024 International Conference on Parallel Architectures and Compilation Techniques},
pages = {91–107},
numpages = {17},
location = {Long Beach, CA, USA},
series = {PACT '24}
}

@inproceedings {tcgnn,
author = {Yuke Wang and Boyuan Feng and Zheng Wang and Guyue Huang and Yufei Ding},
title = {{TC-GNN}: Bridging Sparse {GNN} Computation and Dense Tensor Cores on {GPUs}},
booktitle = {2023 USENIX Annual Technical Conference (USENIX ATC 23)},
year = {2023},
isbn = {978-1-939133-35-9},
address = {Boston, MA},
pages = {149--164},
url = {https://www.usenix.org/conference/atc23/presentation/wang-yuke},
publisher = {USENIX Association},
month = jul
}

@article{bajaj2024graphneuralnetworktraining,
      author={Saurabh Bajaj and Hojae Son and Juelin Liu and Hui Guan and Marco Serafini},
      title={Graph Neural Network Training Systems: A Performance Comparison of Full-Graph and Mini-Batch}, 
year = {2024},
issue_date = {December 2024},
volume = {18},
number = {4},
journal = {Proceedings of the VLDB Endowment},
month = december,
pages = {1196 - 1209},
numpages = {14}
}
\bibliographystyle{mlsys2025}
\appendix
\newpage
\section{Artifact Appendix}

\subsection{Abstract}

The artifact contains the source code of SPA and scripts to run experiments on the baselines to reproduce the results in our paper.
The artifact primarily requires a machine with four NVIDIA V100 GPUs, fully connected via NVLink.
The artifact supports the key results in Table 3, demonstrating SPA's effectiveness in improving end-to-end training time relative to the baselines by eliminating redundant work while minimizing load imbalance and communication costs.

\subsection{Artifact check-list (meta-information)}

\begin{itemize}[itemsep=1pt] 
  \item \textbf{Algorithm:} Graph Neural Networks
  \item {\bf Program: } Pytorch, CUDA and C/C++ Code
  \item {\bf Dataset: } Preprocessed datasets are provided.
  \item {\bf Hardware: } AWS EC2 p3.8xlarge instance consisting of at least 244 GB RAM with 4 NVIDIA V100 GPUs (16GB memory) and GPUs are connected to the CPU with a PCIe 3.0 x 16 bus and with each other via NVLink.
  \item {\bf Metrics: } Execution time
  \item {\bf Output: } Tables and graphs
  \item {\bf Experiments: } We provide a README in the artifact containing instructions to set up, run experiments, and post-process the results to generate tables and figures. The maximum variation is 5\%.
  \item {\bf How much disk space is required (approximately)?: } The artifact requires approximately 300GB to save all the data and graphs.
  \item {\bf How much time is needed to prepare workflow (approximately)?: } We provide a compile script (build.sh) which takes 10 min and download script (download.sh) which takes 50 min.
  \item {\bf How much time is needed to complete experiments (approximately)?: } The total execution time takes approximately 12 hours.
  \item {\bf Publicly available?: } Yes 
  \item {\bf Code licenses (if publicly available)?: } Apache-2.0
  \item \textbf{ Archived (provide DOI)?: } Zenedo DOI will be provided after evaluation. 
\end{itemize}

\subsection{Description}
All computational artifacts are in a single repository, in
which the README contains the detailed instructions to set up and reproduce the figures and tables in the paper:

\url{https://github.com/Juelin-Liu/dgl.git} (Branch: spa-mlsys-ae)

\subsubsection{Hardware dependencies}

By default, our experiments use an AWS EC2
p3.8xlarge instance with 4 NVIDIA V100 GPUs (16GB
memory) and Xeon E5-2686 v4 @ 2.70GHz, with 32 CPU
cores and 244 GB RAM. GPUs are connected to the CPU
with a PCIe 3.0 x 16 bus and to each other via NVLink.
For experiments with 8 GPUs, we use a similar p3.16xlarge instance having 64 CPU cores and 488 GB of RAM.

\subsubsection{Software dependencies}

The artifact uses CUDA v11.8 and
GCC version 11.3 (-O2) to compile all the tested systems. 
We use PyTorch v2.0 as the backend for all the tested systems.
We recommend using Docker to setup the software dependencies and provide instructions for the setup in our README file.

\subsubsection{Datasets}

We provide pre-processed datasets, including the graph topology and partition maps that can be downloaded from Amazon S3. 
Alternatively, you can use the scripts in the repository to generate the prepared
datasets. 
Notice that this would take several days on a single machine to generate all partition maps for the graphs.

\subsection{Installation}
We provide a Dockerfile for setting up the environment for compiling the source code. The instructions for running the Dockerfile are provided in the README. 
We also provide scripts used for downloading the dependencies if you prefer not to use Docker.
Please note to set the branch to spa-mlsys-ae and recursively pull all the submodules, before running the Docker container. 
After installing all the dependencies, and pulling all submodules, simply execute the build.sh script to install Spara into your Conda environment.

\subsection{Experiment workflow}

\begin{itemize}
    \item[$S_1$] The first step is to obtain the input datasets, which include the graph topology data and partition maps.
    The script \textit{download.sh} can be used to download these pre-processed files automatically. 
    \item[$S_2$] After obtaining the prepared datasets, you can run the main experiment by executing the bash script\  \textit{experiment/script/main.sh}. This script runs all the baselines and generates the log file. (Expected time: 120 min, depends on $S_1$.)
    \item[$S_{3}$] Run the script $experiment/partition\_ablation$ to collect the training logs for varying partitioning strategies. (depends $S_1$)
    \item[$S_4$] We postprocess the logs from $S_2$ and $S_3$ to generate Figures 3 using the notebook \textit{$plot/time\_breakdown$}. (depends on $S_2$ and $S_3$).
    \item[$S_5$] Postprocess the logs from $S_2$ and $S_3$ using the notebook \textit{$plot/main$} to generate Table 3 showing the relative performance of spara against the baselines.  (depends on $S_2$ and $S_3$).
    \item[$S_6$]  Run the sampling simulation \textit{$experiment/sample\_main$}, to generate the varying edges computed and features loaded for varying batch sizes (1024,256) and graphs(papers100M, orkut, friendster) to generate the datapoints in Table 1. (Expected time: 30min, depends on $S_1$)
    \item [$S_7$] Run the script
    \textit{$experiment/scripts/ablation.sh$} to run all the ablation experiments on papers100M graph. (depends on $S_1$)
    \item [$S_8$] Post-process the logs generated in the previous step with the Jupyter notebook \textit{$plot/final\_ablation$} (Expected time: 3 hours, depends on $S_6$)
    \item [$S_9$]: Run the script  \textit{$experiment/script/simulate$} to generate the workload characteristics for various partitioning schemes (depends on $S_1$)
    \item [$S_{10}$]: Post process the workloads generated in step $S_8$ with the notebook \textit{$plot/simulation\_plot$} to generate Figures 5. (depends on $S_8$)
\end{itemize}
\subsection{Evaluation and expected result}

The key results of the paper can be demonstrated at $S_5$ in our workflow, where we generate the results in Table 3, demonstrating the effectiveness of spa's split parallelism relative to the baselines. 

\subsection{Notes}

The artifact can be checked for functionality on a machine with four GPUs with or without NVLink connections.  

\subsection{Methodology}

Submission, reviewing and badging methodology:

\begin{itemize}
  \item \url{http://cTuning.org/ae/submission-20190109.html}
  \item \url{http://cTuning.org/ae/reviewing-20190109.html}
  \item \url{https://www.acm.org/publications/policies/artifact-review-badging}
\end{itemize}

\end{document}